\documentclass[11pt,a4paper]{article}
\usepackage[ngerman,english]{babel}
\pdfoutput=1 

\usepackage{jinstpub} 

\usepackage{bm}
\usepackage[usenames,dvipsnames]{xcolor}
\usepackage{comment}
\usepackage[version=4]{mhchem}
\usepackage{siunitx}

\usepackage{lineno}
\DeclareSIUnit\bar{bar}
\DeclareSIUnit\slpm{slpm}
\DeclareSIUnit\permille{\text{\textperthousand}}
\usepackage{tabularx}
\usepackage{ltablex}
\usepackage{array}
\usepackage{enumitem}
\usepackage{booktabs}
\usepackage[font={small},labelfont={bf}]{caption}
\usepackage[font={small}]{subcaption}
\usepackage{blindtext}
\usepackage{nicefrac}
\defineshorthand{"~}{\babelhyphen{nobreak}}
\useshorthands{"}

\newcommand*\diff{\mathop{}\!\mathrm{d}}

\title{Benchmarking the design of the cryogenics system for the underground argon in DarkSide-20k}

\collaboration{\includegraphics[height=17mm]{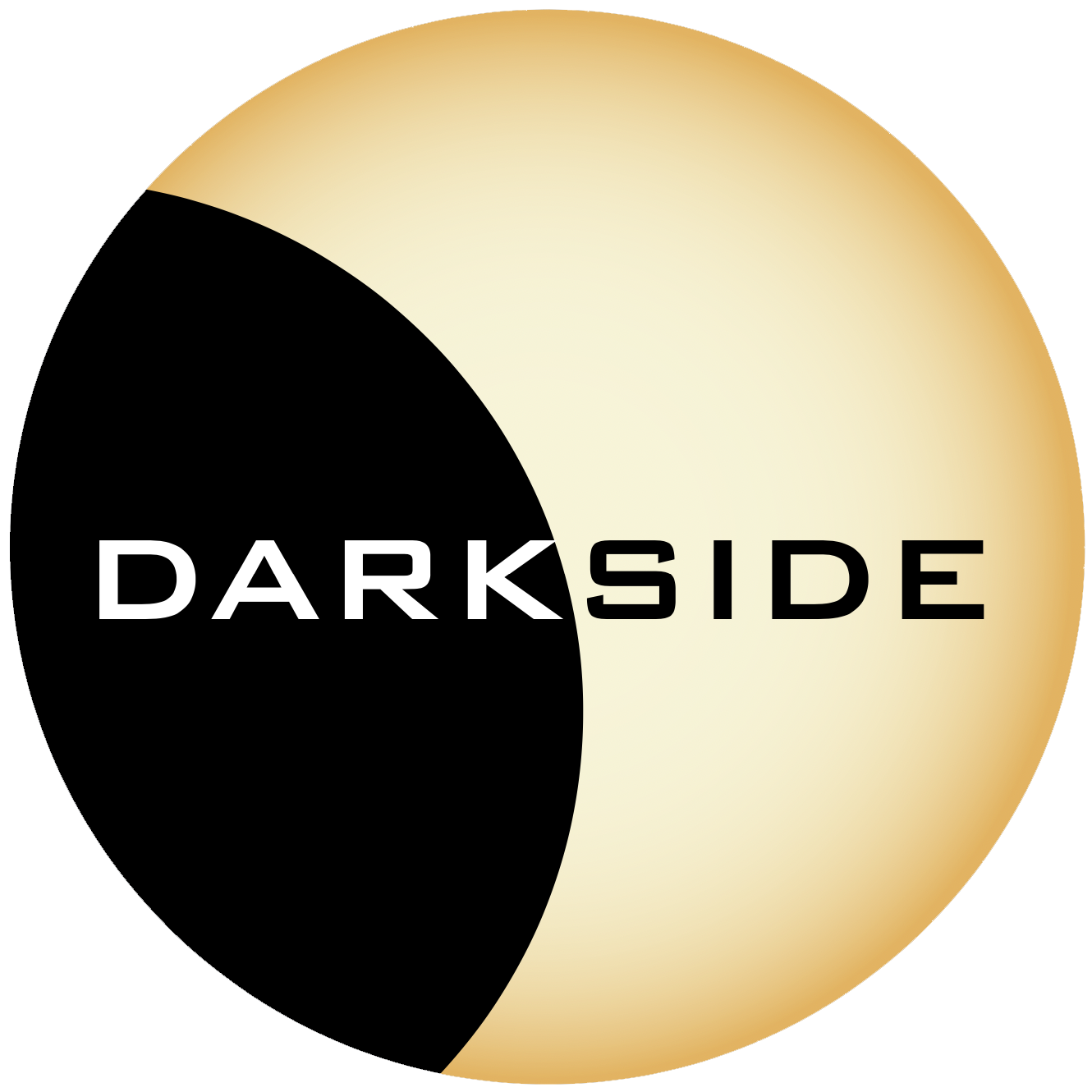}\\[6pt]
The DarkSide-20k Collaboration}

\author[1] {F.~Acerbi,}
\affiliation[1]{Fondazione Bruno Kessler, Povo 38123, Italy}
\author[2]{P.~Adhikari,}
\affiliation[2]{Department of Physics, Carleton University, Ottawa, ON K1S 5B6, Canada}
\author[3,4]{P.~Agnes,}
\affiliation[3]{Gran Sasso Science Institute, L'Aquila 67100, Italy}
\affiliation[4]{INFN Laboratori Nazionali del Gran Sasso, Assergi (AQ) 67100, Italy}
\author[5] {I.~Ahmad,}
\affiliation[5]{AstroCeNT, Nicolaus Copernicus Astronomical Center of the Polish Academy of Sciences, 00-614 Warsaw, Poland}
\author[6,7]{S.~Albergo,}
\affiliation[6]{INFN Catania, Catania 95121, Italy}
\affiliation[7]{Universit\`a of Catania, Catania 95124, Italy}
\author[8]{I.~F.~M.~Albuquerque,}
\affiliation[8]{Instituto de F\'isica, Universidade de S\~ao Paulo, S\~ao Paulo 05508-090, Brazil}
\author[9]{T.~Alexander,}
\affiliation[9]{Pacific Northwest National Laboratory, Richland, WA 99352, USA}
\author[10]{A.~K.~Alton,}
\affiliation[10]{Physics Department, Augustana University, Sioux Falls, SD 57197, USA}
\author[11]{P.~Amaudruz,}
\affiliation[11]{TRIUMF, 4004 Wesbrook Mall, Vancouver, BC V6T 2A3, Canada}
\author[3,4]{M.~Angiolilli,}
\author[12]{E.~Aprile,}
\affiliation[12]{Physics Department, Columbia University, New York, NY 10027, USA}
% ...

\emailAdd{ds-ed@lists.infn.it}

\abstract{
DarkSide-20k~(DS-20k) is a dark matter detection experiment under construction at the Laboratori Nazionali del Gran Sasso~(LNGS) in Italy. It utilises $\SI{\sim 100}{t}$ of low radioactivity argon from an underground source~(UAr) in its inner detector, with half serving as target in a dual-phase time projection chamber~(TPC).
The UAr cryogenics system must maintain stable thermodynamic conditions throughout the experiment's lifetime of over $\SI{10}{years}$. Continuous removal of impurities and radon from the UAr is essential for maximising signal yield and mitigating background. 
We are developing an efficient and powerful cryogenics system with a gas purification loop with a target circulation rate of $\SI{1000}{slpm}$. Central to its design is a condenser operated with liquid nitrogen which is paired with a gas heat exchanger cascade, delivering a combined cooling power of more than $\SI{8}{kW}$. 
Here we present the design choices in view of the DS-20k requirements, in particular the condenser's working principle and the cooling control, and we show test results obtained with a dedicated benchmarking platform at CERN and LNGS. 
We find that the thermal efficiency of the recirculation loop, defined in terms of nitrogen consumption per argon flow rate, is $\SI{95}{\%}$ and the pressure in the test cryostat can be maintained within $\pm(\numrange{0.1}{0.2})\,\si{mbar}$. We further detail a 5-day cool-down procedure of the test cryostat, maintaining a cooling rate typically within $\SI{-2}{K/h}$, as required for the DS-20k inner detector. Additionally, we assess the circuit's flow resistance, and the heat transfer capabilities of two heat exchanger geometries for argon phase change, used to provide gas for recirculation. 
We conclude by discussing how our findings influence the finalisation of the system design, including necessary modifications to meet requirements and ongoing testing activities.}


\arxivnumber{2408.14071} 

\begin{document}


\maketitle
\flushbottom

\section{Introduction}
\label{sec:Introduction}

DarkSide"~20k~(DS"~20k) is a direct dark matter search experiment, currently under construction underground in Hall~C at the Laboratori Nazionali del Gran Sasso~(LNGS) in Italy~\cite{DarkSide-20k:2017zyg}. Compared to its predecessor experiment DarkSide"~50~(DS"~50)~\cite{DarkSide:2018kuk}, DS"~20k will substantially improve the sensitivity to the elastic scattering of weakly interacting massive particles~(WIMPs) off nuclei. For a fiducial volume exposure of $\SI{200}{t \cdot y}$, a spin-independent cross-section exclusion limit (at $\SI{90}{\%}$ C.L.) of $\SI{6.3e-48}{cm^2}$ is projected for $\SI{1}{TeV/c^2}$ WIMPs.

At its core, DS"~20k deploys a dual-phase (liquid/vapour) argon time projection chamber (TPC). Dual-phase TPCs employing liquefied noble gases for dark matter searches use photosensors to detect scintillation light produced by particle interactions with target atoms. While the primary scintillation light signal~(S1) produced in liquid is directly detected, the ionisation charge is drifted upward by a vertical electric field, and extracted into the gas argon~(GAr) phase, where the electrons stimulate an electroluminescence signal~(S2). Based on these signals, the recoil energy and interaction site can be reconstructed. The use of liquid argon~(LAr) allows the powerful discrimination of background electronic recoil events from signal-like nuclear recoils by means of the primary scintillation signal temporal topology (pulse-shape discrimination).

\begin{figure}[t]
\centering
\hspace{2.75cm}
\includegraphics[width=0.618 
\textwidth]{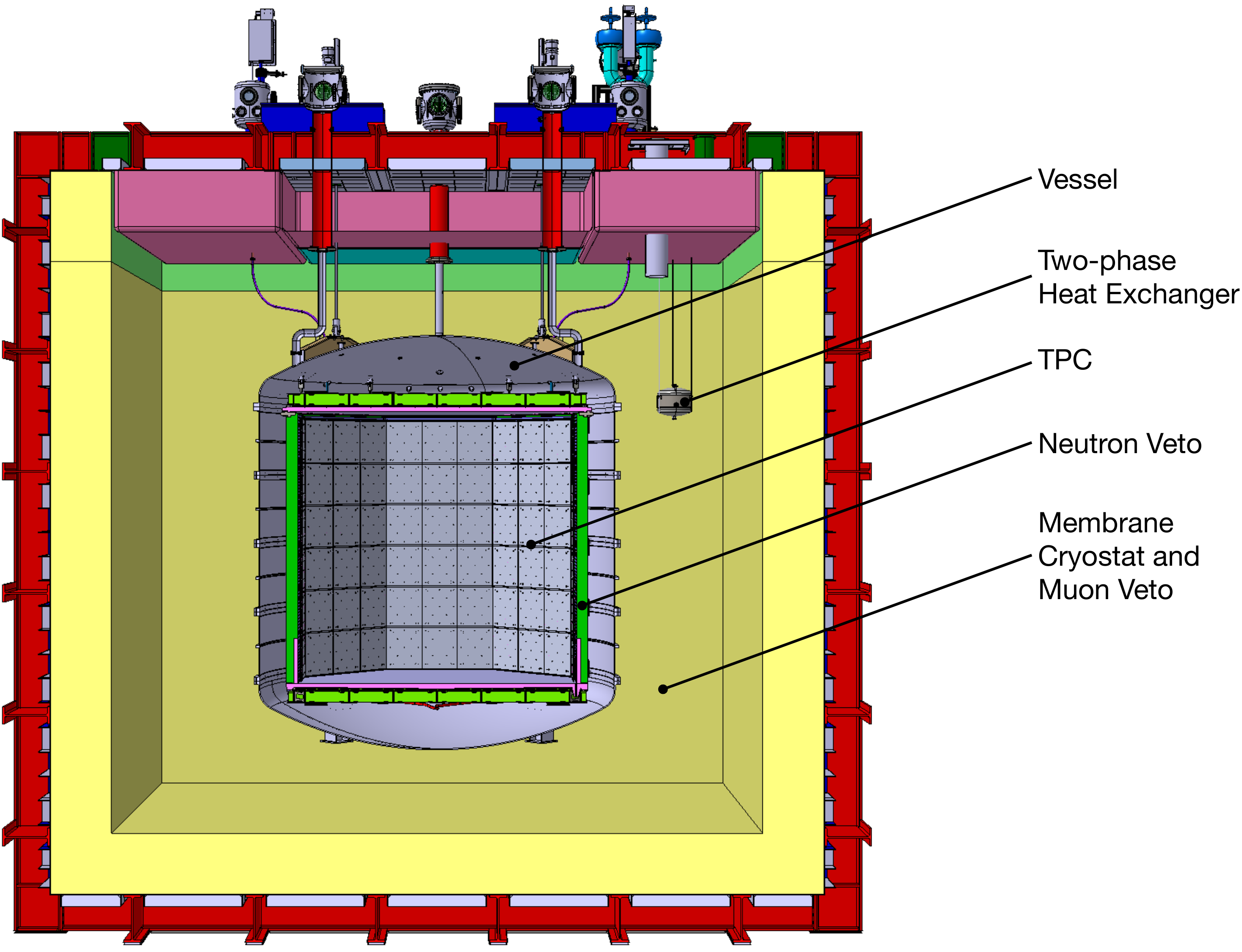}
\caption[]{Schematic overview of the DarkSide-20k detector. The UAr is contained inside the vessel, the AAr inside the membrane cryostat.}
\label{fig:DS-20kSchematicOverview}
\end{figure}

The DS"~20k detector has a nested design, as shown is figure~\ref{fig:DS-20kSchematicOverview}. The innermost subsystem is the TPC, surrounded by an active neutron detector, constituting together the inner detector. The inner detector is in turn surrounded by the outer veto detector, which is bounded by a ProtoDUNE-like membrane cryostat~\cite{DUNE:2021hwx}. The inner and outer detectors are separated by a stainless steel vessel creating two separate volumes inside the cryostat.

The DS"~20k TPC has an octagonal shape and its structure is made of acrylic. The signals are read by two planes of custom-made cryogenic silicon photomultipliers~(SiPMs) above and below the TPC. 
The vessel containing the inner detector is filled with $\SI{\sim 100}{t}$ of low-radioactivity argon, extracted from an underground source~\cite{Acosta-Kane:2007you,Back:2012pg}. Half of this quantity is contained inside the active region of the TPC. As demonstrated by DS"~50, the content of the $\beta$-emitting isotope \ce{^{39}Ar} in underground argon~(UAr), extracted from a CO$_2$ well in Colorado, is more than a factor 1400 lower than in regular argon of atmospheric origin~(AAr)~\cite{DarkSide:2014,DarkSide:2015}.
The outer veto detector has a SiPM readout and deploys $\SI{\sim 650}{t}$ of AAr, which are contained in the cryostat. 

The DS-20k design requires two separate, but not independent, cryogenic systems: one for the AAr in the outer detector, and another one for the high-purity UAr, which will be the topic of this manuscript. The UAr cryogenics system is responsible for the initial cool-down and filling of the inner detector with UAr, the compensation of heat loads from detector instrumentation and the overall thermodynamical stability of the detector. It further allows the formation of a gas pocket for S2 signal generation, the continuous removal of electronegative impurities (which reduce the S1~and S2~signal yield and change the shape of the S1~signal) and of \ce{^{222}Rn} (one of the major background sources) from the UAr as well as the deployment of internal calibration sources. To reach its sensitivity goals, DS"~20k is designed to operate for a minimum of $\SI{10}{years}$, posing stringent stability and reliability requirements on all its systems, in particular on the cryogenics system that handles the UAr. This manuscript presents the conceptual design and technical solutions of the UAr cryogenics system fulfilling these requirements. It further provides benchmarking data that was acquired with a dedicated test bed at CERN and LNGS containing core elements of the final DS"~20k cryogenics system.

The paper is structured as follows: section~\ref{sec:RequirementsConceptDesign} provides a discussion of the system's requirements and a conceptual high-level system description with purpose, functionality and design choice reasoning. It further details the interplay of the components and touches upon the system's operating modes. In section~\ref{sec:TechnicalDesign}, we provide a technical description and design details of the various components of the system that underwent thorough testing, along with an overview of the testing facilities. The test results and reached performance benchmarks can be found in section~\ref{sec:CommissioningBenchmarking}. We discuss the implications of the tests for the DS"~20k system design, provide a brief outlook on ongoing testing activities, and summarise our work in section~\ref{sec:DS-20kOutlook}.
Throughout this work, we utilise fluid properties from the NIST Standard Reference Database~\cite{NIST}. We denote absolute, gauge and differential pressure units by appending a, g and d, and use $\si{bara}$, $\si{barg}$, $\si{bard}$, respectively with their appropriate unit prefixes.
\section{System requirements and conceptual design}
\label{sec:RequirementsConceptDesign}

The system is designed upon the successful cryogenics of DS"~50~\cite{DarkSide:2014,DarkSide-50:2023nes}, which operated reliably for $\SI{5}{years}$. Based on the experience gained with this system we can identify the following list of key requirements on the DS"~20k UAr cryogenics: 
\begin{itemize}[itemsep=0pt]
\item{A gas filling and recirculation flow rate of $\SI{1000}{slpm}$: this requirement ensures a timely filling and turnover of the entire UAr mass of $\SI{100}{t}$ within $\SI{40}{days}$. Thus, given that the concentration of electronegative impurities initially reduces by a factor of $1/e$ per turnover (see reference~\cite{Plante:2022khm} for a modelling of liquid xenon purification), the target purity (see purity requirement below) can be reached within the first year of operation, assuming an initial impurity concentration of the order of $\SI{1}{ppm}$. It is expected that the recirculation flow can be reduced after an initial purification phase. This requirement sets demands on circuit flow resistance, recirculation pump performance, heat recovery efficiency and cooling power.}
\item{Cooling power of minimum $\SI{8}{kW}$: this is the power needed to cool and liquefy argon at the design flow rate, at a pressure of $\SI{1}{bara}$ and when starting at room temperature. Filling is the phase with the highest cooling power demand. While recirculating during operation, heat must be recovered using heat exchange for energy use efficiency (see next point). The detector instrumentation accounts for $\SIrange{2.0}{2.5}{kW}$ static heat load. Heat loads from radiative and convective heat transfer between the cryogenic components that are located outside of the cryostat and the environment at ambient temperature can be assumed negligible here due to the use of multi-layer and vacuum insulation.}
\item{High heat recovery efficiency: due to the experiment's anticipated duration of more than $\SI{10}{years}$, efficient energy use is essential. In particular, an efficient heat transfer between the argon supply and return flow of the recirculation loop must be ensured to minimise the dynamic heat load.}
\item{Long-term pressure stability: the pressure in the UAr vessel containing the TPC is required to be stable within a root mean square~(RMS) of $\SI{1}{mbar}$. This requirement ensures the stability and resolution of the S2 signal to within $\SI{0.1}{\%}$. DS"~50 has demonstrated stable operations with pressure fluctuations with a span of $\SI{0.35}{mbar}$. Pressure and temperature fluctuations at this level were predicted to influence the electroluminescence yield below a level of $10^{-4}$, consistent with observations~\cite{DarkSide-50:2023nes}.} 
\item{Low impurity level: an oxygen-equivalent impurity level of less than $\SI{0.06}{ppb}$ (parts per billion) is required for a minimum free electron lifetime of $\SI{5}{ms}$ at $\SI{200}{V/cm}$ drift field strength and $\SI{3.5}{m}$ drift distance~\cite{Bakale1976,Li:2022pfu}.}
\item{Leak-tightness: the argon-wetted gas components are required to have a leak rate of lower than $\SI{e-8}{mbar\cdot l \cdot s^{-1}}$ helium-equivalent to avoid contamination with air and radon background.}
\item{Enable all operating modes: the system must allow to perform all operations that are required to run DS"~20k. These include the preparatory evacuation of the process volumes and insulation vacua, the cooling of the detector, the filling of the detector from both LAr and GAr sources, the normal operation with recirculation and purification, the sampling of UAr for purity analysis, the introduction of distributed internal calibration sources, and the recovery of the UAr.}
\item{Power failure immunity: to avoid UAr loss, cooling must be assured in the event of a power blackout. The underground location of the system further necessitates safe and reliable sustainment of stable operating conditions for an extended period in case of access restriction. This requirement includes the failure of uninterruptible power supplies.}
\item{Redundancy, reliability and maintainability: due to the long projected runtime of more than $\SI{10}{years}$ the system must be designed robust and durable while enabling the exchange or maintenance of individual components during ongoing operation.}
\end{itemize}

\begin{figure}[t]
\centering
\includegraphics[width=\textwidth]{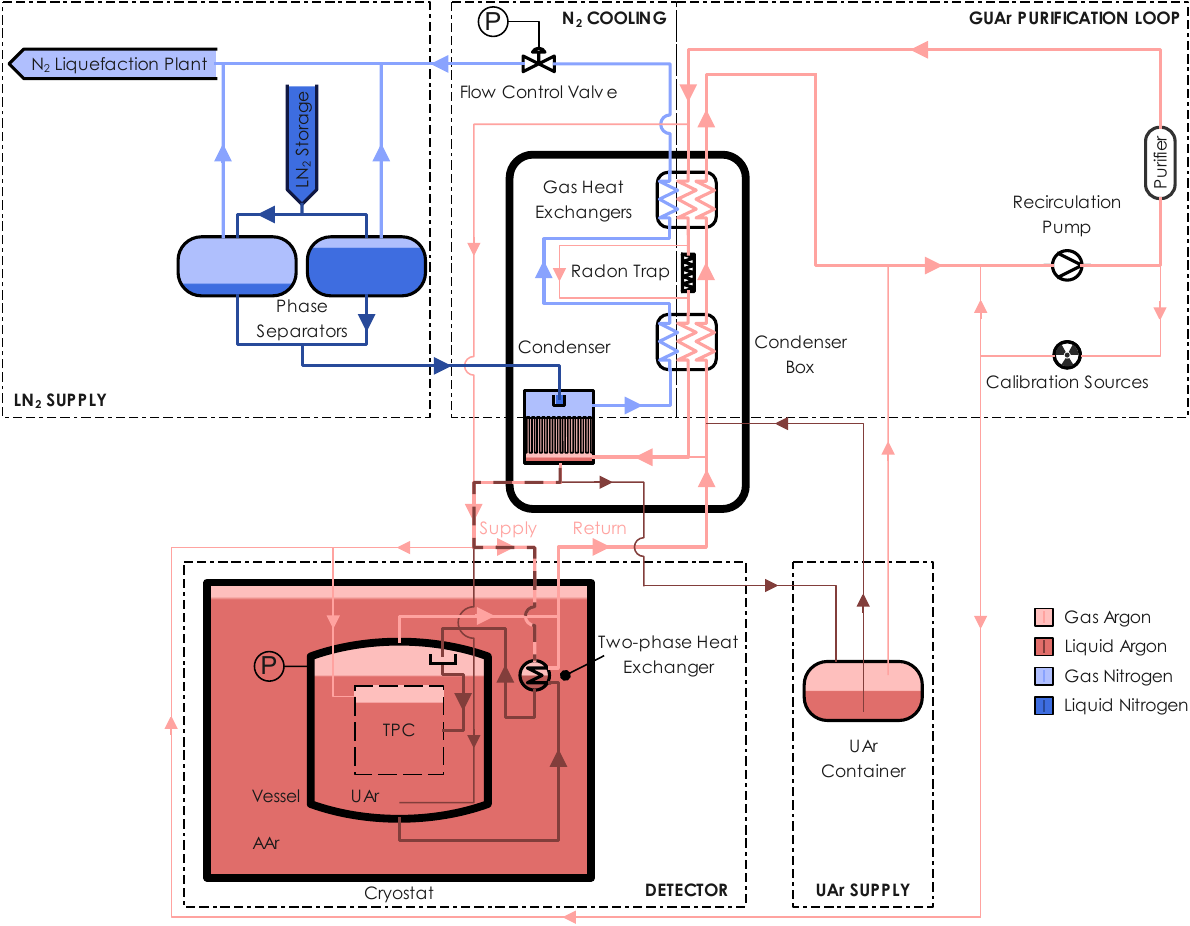}
\caption[]{Conceptual P\&ID of the DS"~20k UAr cryogenics system. The paths active during normal operation are denoted by thick lines, while alternative paths used in other operating modes are indicated by thin lines. By~P we denote the pressure transducer based on whose signal the cooling is controlled.}
\label{fig:DS-20kConceptualPID}
\end{figure}

Next, we provide a high-level description of the system's conceptual design that was developed to meet this set of requirements. Figure~\ref{fig:DS-20kConceptualPID} shows the corresponding piping and instrumentation diagram~(P\&ID)\footnote{An engineering P\&ID of the full DS"~20k UAr cryogenics system can be found in reference~\cite{DS20kPID}.}, which depicts the main building blocks of the system: detector, GAr purification loop, UAr supply, liquid nitrogen~(LN$_2$) supply and nitrogen~(N$_2$) cooling. 

As explained in the preceding section, the detector features two separate argon volumes for the AAr and the UAr. The bare vessel containing the UAr is immersed in the AAr without insulation, providing a large contact surface for a potential heat transfer. While the AAr is served by a separate cryogenics system, both volumes are thermally coupled. Effectively, the operating conditions of the vessel and thus of the UAr cryogenics system are dictated by the conditions of the AAr in the cryostat. On the one hand, it would require an immense cooling power to reduce the temperature of the UAr volume below the temperature of the AAr. On the other hand, it must be avoided that the AAr cryogenics system is stressed by any significant, in particular varying, heat load from the UAr reservoir. The ullage pressures of the cryostat and the vessel must thus be equal in equilibrium to eliminate a temperature gradient across the vessel. The operating pressure is chosen above the ambient pressure in the underground cavern to reduce the risk of air leakage into the system and its choice depends on details of the AAr cryogenics system and the cryostat pressure rating. While the exact setpoint has not been fixed yet, a value in the range $\SIrange{1040}{1075}{mbara}$ is considered. From ProtoDUNE"~SP operations, it is clear that the temperature gradients inside the bulk of the LAr are low, of the order of $\SIrange{1}{10}{mK}$~\cite{DUNE:2021hwx}. Hence, we do not expect temperature gradients over the vessel's surface. 
 
To achieve the required recirculation flow rate, LAr must be efficiently boiled off. The height to which LAr can be extracted by means of underpressure, to subsequently perform the two-phase heat exchange, is limited by the difference between the gas ullage pressure and the triple point pressure. For the range of ullage pressures given above, this height difference is $\SIrange{2.5}{2.8}{m}$, which is insufficient for the experiment's dimensions. Thus, if penetrations through the cryostat walls are best avoided to minimise associated risks, the phase change has to take place inside the membrane cryostat. To this end, a two-phase heat exchanger is placed at the depth of the vessel's liquid-gas interface inside the AAr. The supply side of the two-phase heat exchanger is downstream of the purification loop and thus at a higher pressure than the return side. Hence, the supplied argon has a higher saturation temperature than the returned argon. This temperature offset allows for a phase transition on both sides. While passing through the heat exchanger, the supplied GAr transfers heat to the return side and undergoes liquefaction. In response, the LAr on the return side evaporates, which provides GAr for recirculation. Since the heat exchanger is not insulated, there is a non-zero heat transfer from the UAr supply side to the AAr bath. The power of this process is estimated to be $\SIrange{450}{650}{W}$ for the considered range of differential pressures and the multi-tube design presented in section~\ref{subsec:TwoPhaseHE}. This heat load can be handled by the AAr cryogenics system. If this additional cooling power to the UAr cryogenics system is not desired, such as it is the case when the TPC instrumentation is not operational, either the pressure drop over the heat exchanger can be reduced by decreasing the compressor output, or the two-phase heat exchanger can be bypassed. The liquid UAr is injected into the TPC's active volume at designated locations and in proximity of the photoelectronics to induce convection for heat dissipation. The gas pocket in the TPC can be created and maintained by means of a direct GAr supply through a GAr line connected to the top of the TPC or by evaporating LAr in situ with electric resistive heater elements (not shown in the P\&ID). 

GAr and LAr supply and return process lines are routed between the vessel and purification loop to allow for filling, recirculation and recovery of UAr. The closed purification loop consists of the following components: gas-gas heat exchangers between supply and return side maximise energy use efficiency; a condenser compensates heat loads of the system by removing thermal energy through argon phase change; one or several recirculation pump(s) create the required pressure gradient to drive a flow; an injection system allows the deployment of gaseous calibration sources; a zirconium-based hot getter removes electronegative molecules\footnote{The model PS5 from Entegris MegaTorr\textsuperscript{\textregistered} removes impurities with standard specification below the $\SI{1}{ppb}$ level~\cite{Getter}.}; and a radon trap removes \ce{^{222}Rn}. The latter is situated before the last, in the direction of the supply flow, and coldest gas-gas heat exchanger in the condenser box. This ensures that the trap is operated at low temperature while preventing LAr from entering the trap during operations. The breakthrough time of radon, and thus the adsorption efficiency, is significantly enhanced at low temperatures, which guarantees that \ce{^{222}Rn} has decayed before passing the trap~\cite{Pushkin:2018wdl}. While direct liquid purification~\cite{DUNE:2021hwx,XENON:2024wpa} has been considered as a potential design option, in addition to a gas purification system, sole purification of GAr was chosen for its simplicity and reliability, and in lack of an established method to remove nitrogen from LAr efficiently. This choice avoids the presence of piping connected to the side of the cryostat that carries UAr and the use of a pump for cryogenic liquids, and reduces associated risks. Extrapolating from the DS"~50 experience~\cite{DarkSide-50:2023nes}, the achievable turnover of $\SI{40}{days}$ with GAr purification is expected to be sufficient for the purity requirements. 

Gaseous internal calibration sources can either follow the standard recirculation path, appropriate for the deployment of \ce{^{83\text{m}}Kr} with a half-life of $\SI{1.83}{h}$~\cite{McCutchan:2015vcl}, or bypass the purifier and condenser box and be directly injected into the gas pocket of the TPC. The latter procedure may be adopted for quickly decaying sources such as \ce{^{220}Rn} with a half-life of $\SI{56}{s}$~\cite{ELLIS1976341}.

For filling and recuperation, the system deploys $\SI{20}{ft}$~ISO shipping and storage containers for cryogenic liquids, which are used for transporting UAr from its extraction plant URANIA in Colorado to the distillation plant ARIA in Sardinia~\cite{DarkSide-20k:2023grj} and finally to DS"~20k at LNGS. These containers feature a separate LN$_2$-based condenser and a heater. The cryogenics system can accept boil-off GAr as well as LAr. Both filling paths ensure that the argon is cycled through the purification loop before reaching the detector. The liquid filling path includes a phase change in the first heat exchanger stage of the condenser box and a recondensation after purification. Recovery is possible via a path from the condenser to the UAr container. To evaporate LAr from the vessel, heat can be introduced via a bypass line around the condenser box, which connects the purifier outlet directly with the detector inlet. 

Cooling is provided by the condenser, which is served with LN$_2$ from an external storage. Two decoupled $\SI{500}{L}$ phase separators, used alternately, ensure a continuous supply of LN$_2$ at constant pressure. While the phase separator open to the UAr cryogenics system is in use and stable, the other one will be isolated and refilled from the LN$_2$ storage\footnote{Unless compensated by an active pressure adjustment, there will be a remaining pressure step in the condenser of the order of $\SI{10}{mbard}$ when switching from one tank to the other due to the difference in liquid level.}. After latent heat exchange inside the condenser, boil-off gas nitrogen~(GN$_2$) is further warmed by heat transfer from the supply GAr, pre-cooling the latter, and thus increasing the efficiency\footnote{Here we refer to the efficiency of the UAr cryogenics system excluding the external N$_2$ liquefaction plant.} and cooling power of the system. The cooling power is regulated by adjusting the GN$_2$ flow leaving the condenser box, depending on the vessel pressure. The warm GN$_2$ is recycled in an external liquefaction plant. It is important to note that the system is operated with LN$_2$ which is only slightly pressurised to not more than $\SI{1.5}{bara}$, as necessary for high GN$_2$ flow rates at more than $\SI{1000}{slpm}$, but with a nitrogen saturation point below the triple point temperature of argon. Compared to a system designed to use pressurised LN$_2$ above $\SI{2.1}{bara}$ to operate with an LN$_2$ temperature above the triple point of argon, this choice reduces the risk associated with the loss of pressurisation. Such a system would not be prepared to operate at an LN$_2$ temperature below the triple point of argon and thus solidify argon on the condenser in case LN$_2$ overpressure is lost. The system presented here instead uses a sophisticated self-regulating dosing mechanism in the condenser, which is explained in detail together with the cooling control in the subsequent section. An additional path directly connects the detector and the condenser via the return path, bypassing the purification loop. This path is assumed during a power failure event and assures uninterrupted cooling of the vessel. The pull of the cold condenser and the liquid column downstream of it maintain a continuous and directed flow, keeping the vessel's pressure constant through liquefaction of boil-off gas.

To verify the correct functioning of certain system aspects, to benchmark individual elements or to inform the design and dimensioning of unfinalised components, we have built a test bed that contains the core components to be deployed in DS"~20k. This is the topic of the next sections.
\section{Technical design of the test bed}
\label{sec:TechnicalDesign}

In the following, we describe the tested configuration, starting with the P\&ID, and detail the technical design of the components used in the test bed. On the one hand, a good fraction of the parts shown here are integral components of the DS"~20k UAr cryogenics system, which were introduced conceptually in the preceding section and which will be installed in the same or in a slightly modified configuration in the final system. On the other hand, the test bed contains certain placeholder components whose benchmarking informs the design choice of the final system for DS"~20k. Among those are two candidate two-phase heat exchanger geometries. Finally, we provide an overview of the monitoring software and framework used in the tests and introduce the test facilities.  

\subsection{Piping and instrumentation diagram}
\label{subsec:P&ID}

\begin{figure}
\centering
\includegraphics[width=\textwidth]{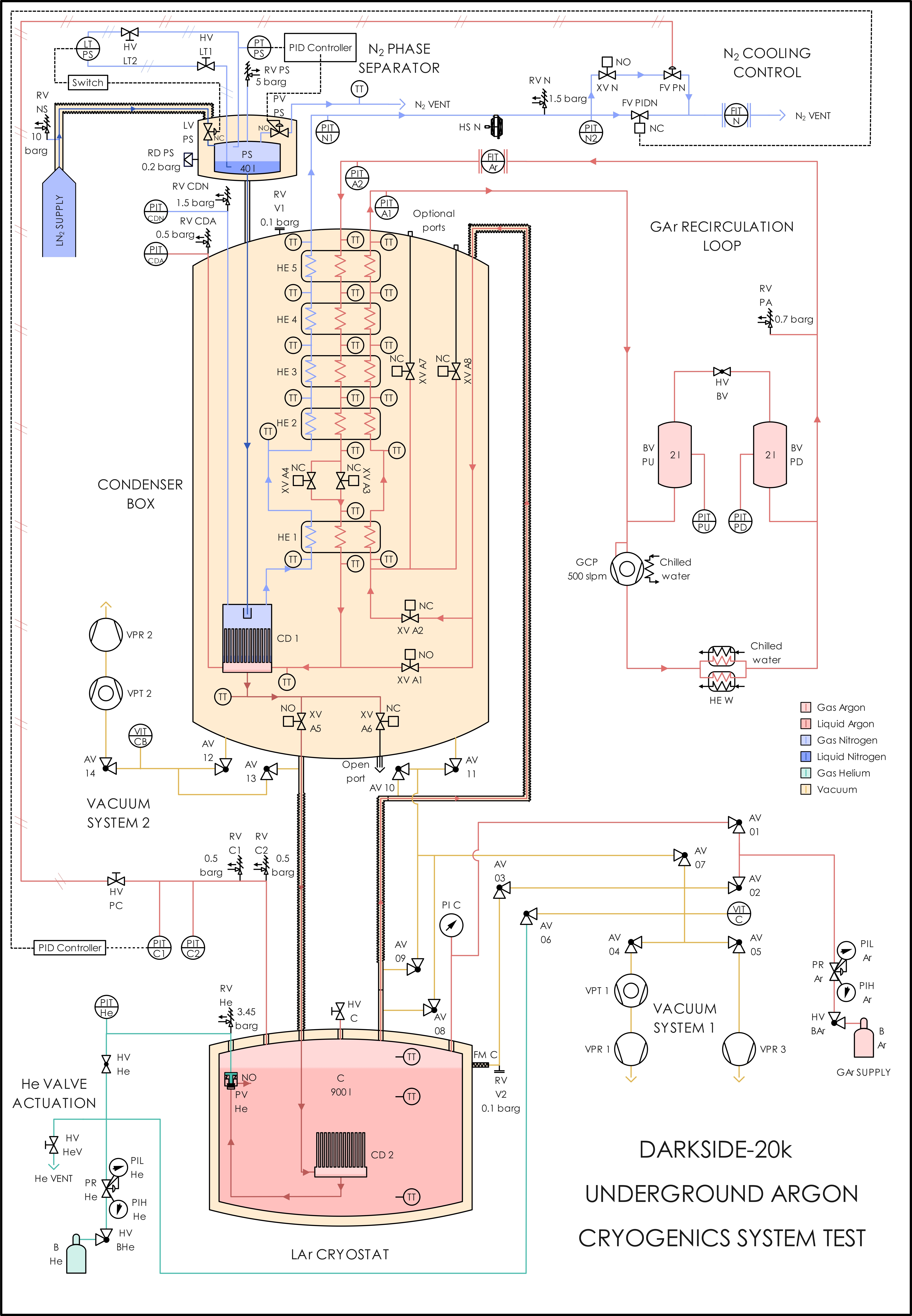}
\caption[]{P\&ID of the UAr cryogenics system test bed. See appendix~\ref{app:PIDSymbols} for a legend of the symbols.}
\label{fig:UArCryoTestPID}
\end{figure}

Figure~\ref{fig:UArCryoTestPID} shows the P\&ID of the test bed. A complete legend of the P\&ID components can be found using their tags in appendix~\ref{app:PIDSymbols}. The operational procedures for the various operating modes of the system can be found in reference~\cite{OpProcedures}. In the P\&ID, the following main components can be identified: the condenser box containing the argon condenser~(CD~1) and a gas heat exchanger cascade~(HE~1--5), a double-walled test cryostat~(C) with an internal volume of $\SI{900}{L}$ containing a two-phase heat exchanger~(CD~2), the GAr recirculation loop featuring a compressor~(GCP), the LN$_{2}$ supply with a $\SI{40}{L}$ phase separator~(PS)\footnote{This is the size of the phase separator used at LNGS. It was chosen as appropriate for the test but is not comparable to the system that will be used in DS"~20k.}, the N$_{2}$ cooling control with control valves~(FV~PIDN, FV~PN), and vacuum systems.

The cryostat is connected downstream of the condenser through flexible transfer lines with an outer diameter~(OD) of $\SI{1}{in}$ and with vacuum- and multi-layer-insulation. Before entering the cryostat volume, the supplied argon passes through a two-phase heat exchanger and a pneumatic valve~(PV~He). Through the pneumatic valve, actuated with pressurised helium from an external gas bottle, a pressure difference between the supply line and the cryostat can be created. This impacts the heat transfer in the two-phase heat exchanger. The valve's outlet, which is submerged in LAr, is viewed by a camera (not shown in the P\&ID), allowing the operator to visually confirm whether solely LAr or a GAr-LAr mixture enters the cryostat. A gas return transfer line brings the cold boiled-off GAr from the cryostat to an outer port of the three-flow heat exchanger cascade in the condenser box, where it is warmed up. The flow is driven by a compressor and measured by a mass flow meter~(FIT~Ar) in the GAr recirculation loop. Downstream of the warm part of the loop, the GAr enters the centre port of the heat exchanger cascade, transferring thermal energy to the GAr return side and the boil-off GN$_{2}$ before arriving at the condenser. The condenser receives LN$_{2}$ from a reservoir located on top of the condenser box, acting as GN$_{2}$-LN$_{2}$ phase separator. The phase separator, constructed by Criotec~\cite{Criotec}, features a control loop for the liquid level and the gas pressure with corresponding control valves as actuators~(LV~PS and PV~PS). The cooling power of the system is determined by the GN$_{2}$ flow out of the condenser, which is regulated by pneumatically actuated valves, that use the cryostat pressure as their input. A gas system allows for the filling of the process media (argon and helium) and evacuation of the wetted and insulation volumes.

Comparing this test setup with the DS"~20k system described in section~\ref{sec:RequirementsConceptDesign}, it is clear that the test setup comprises only the main elements of the final design, with the goal of benchmarking. These are the condenser box, without the radon trap, and the cooling control unit that adjusts the GN$_2$ flow based on a pressure signal. The GAr recirculation loop does not contain a purifier and is in fact the simplest possible, with the goal of testing a compressor candidate for DS"~20k. As demonstrated below, two different small two-phase heat exchanger geometries were deployed in the cryostat, with the goal of understanding the heat transfer in order to scale it appropriately for DS"~20k.

\subsection{Condenser box}
\label{subsec:CondenserBox}

\begin{figure}[t]
\centering
\begin{subfigure}[c]{0.48\textwidth}
\centering
\includegraphics*[height=0.35\textheight]{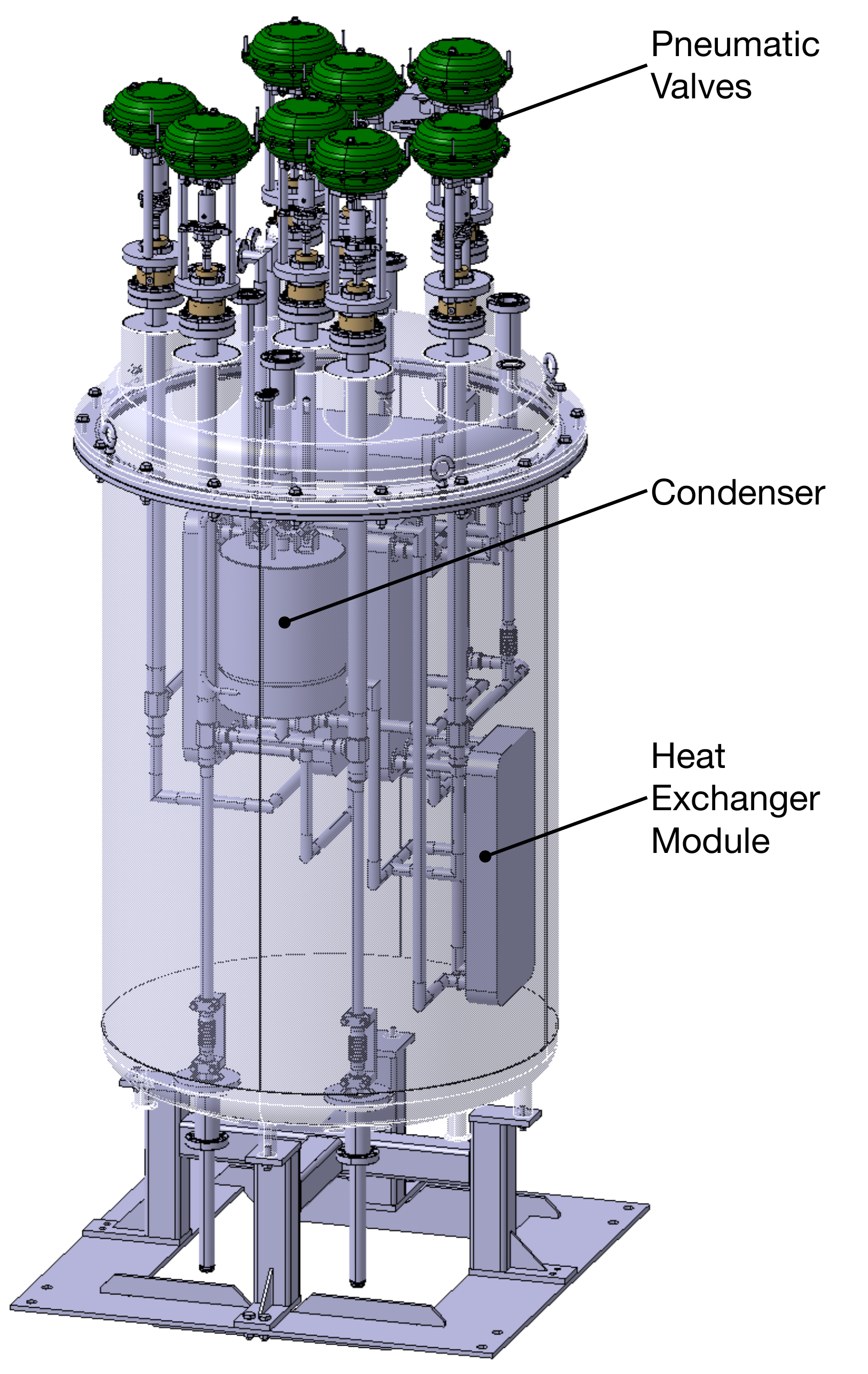}
\end{subfigure}
\quad
\begin{subfigure}[c]{0.48\textwidth}
\centering
\includegraphics*[height=0.35\textheight]{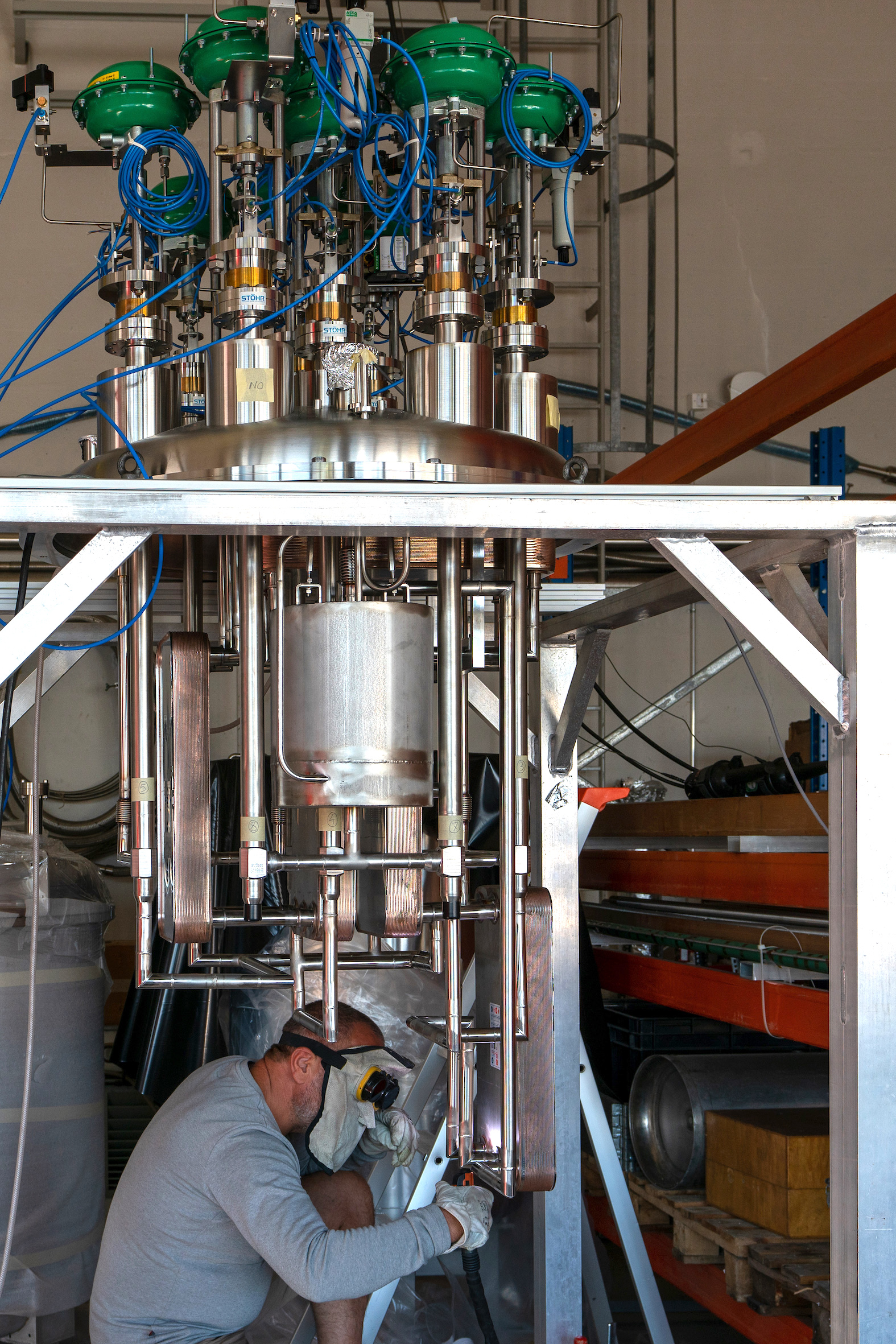}
\end{subfigure}
\caption[]{Condenser box. (Left): CAD model of the condenser box with inside view. (Right): Picture of the construction.}
\label{fig:CondenserBox}
\end{figure}

The condenser box is a vacuum vessel containing the argon condenser~(CD~1), a gas heat exchanger cascade~(HE~1--5), 8~cryogenic valves and the interconnecting $\SI{1}{in}$~OD stainless steel tubes with $\SI{0.065}{in}$ wall thickness (see figure~\ref{fig:CondenserBox}), designed to reach the flow rate and efficiency goals outlined in section~\ref{sec:RequirementsConceptDesign}. The assembly has overall dimensions, including the support structure and the cryogenic valves, of $\SI{\sim 1}{m}$ in diameter and $\SI{\sim 2.5}{m}$ in height with a total mass of $\SI{\sim 570}{kg}$. The condenser box is rated for a maximum pressure of $\SI{0.5}{barg}$ ($\SI{1.5}{barg}$) on the argon (nitrogen) side, and complies with the 2014/68/EU ``Pressure Equipment Directive''~\cite{Condenser}. 

The stainless steel argon condenser is a shell-and-tube heat exchanger with outer dimensions of $\SI{10}{in}\times \SI{12.5}{in}$ (diameter $\times$ height), consisting of 127~tubes with $\nicefrac{1}{2}\,\si{in}$~OD, a wall thickness of $\SI{0.049}{in}$ and a length of $\SI{7}{in}$. The tubes are closed on top and welded at the bottom end to a base-plate, arranged in a hexagonal closely-packed pattern with a pitch of $\nicefrac{3}{4}\,\si{in}$ (cf.~identical design of the two-phase argon heat exchanger in section~\ref{subsec:TwoPhaseHE}). The selection of tube dimensions is informed by heat transfer measurements conducted on different tube geometries. The base-plate separates the nitrogen volume on the top from the argon volume on the bottom. The so-called \textit{chicken feeder}, mounted at the end of the LN$_{2}$ delivery tube passively doses the inflow into the condenser. The chicken feeder consists of a stainless steel bucket in which the LN$_{2}$ tube ends loosely. This provides some backpressure to the LN$_{2}$ line given by the hydrostatic pressure of the depth of the tube's end with respect to the bucket's lip. If the pressure on the nitrogen side of the condenser decreases, due to the opening of the gas vent line, LN$_{2}$ overflows from the bucket and falls into the condenser. This system ensures that the condenser is not flooded with LN$_{2}$ when standard operating procedures are followed and maintains a flow of LN$_{2}$ on an as-needed basis. The chicken feeder doses the inflow such that only a small quantity of LN$_2$ pools on top of the base-plate, establishing a vertical temperature gradient along the tube's length and keeping the temperature of the inner tube surfaces above the triple point of argon. This way, the argon on the other side of the tubes does not solidify under normal operation conditions, even though the nitrogen is not pressurised to a point at which its saturation temperature is higher than the melting point of argon. Instead, the argon condenses on the inside of the tubes, runs down the walls in a thin film and then drops through the transfer line into the cryostat by gravity. The nitrogen undergoes its phase change entirely inside the condenser and the required latent heat is provided by the argon on the other side. The cooling power of the cold GN$_{2}$ is subsequently recycled as described below. 

In order to reach the design gas recirculation flow at minimal LN$_{2}$ consumption, the heat exchanger cascade includes 5~copper-brazed stainless steel dual-circuit parallel-plate heat exchangers (model K215D-22 from KAORI~\cite{KAORI}) with an individual exchange surface of $\SI{2.43}{m^2}$. 
These heat exchangers ensure a highly efficient heat transfer between the three sides, pre-cooling the supply GAr, and warming up the GN$_{2}$ and the return GAr close to ambient temperature. The heat exchangers feature 5~channels each for the two \textit{refrigerant} circuits, in our application the GN$_{2}$ and GAr return, and 11~channels for the \textit{fluid} circuit, in our case the GAr supply. We discuss the impact of this asymmetry between the two argon sides in section~\ref{subsec:CompressorPerformanceCircuitResistance}. 

The total cooling power of the system, i.e.~the combined cooling power of the condenser and the heat exchanger chain, is given by the total change of the nitrogen enthalpy. In an isobaric process, the change of the nitrogen enthalpy is the sum of the nitrogen latent heat transfer during phase change and the enthalpy change related to the increase of the temperature of the gas based on its heat capacity. While the former takes place in the condenser, the latter happens in the heat exchanger chain. The latent heat transfer contributes $\SI{\sim 50}{\%}$ to the total cooling power, as can be seen from the isobaric enthalpy-temperature curve of nitrogen~\cite{NIST}. Thus, to reach the cooling power requirement of minimum $\SI{8}{kW}$ (cf.~section~\ref{sec:RequirementsConceptDesign}), the condenser must allow for a latent heat transfer of minimum $\SI{4}{kW}$. It was determined experimentally with a similar condenser assembly using pressurised argon, that a heat transfer of up to $\SI{9}{W/in}$ of tube can be achieved. For the condenser deployed here, this measurement predicts a cooling power of up to $\SI{8}{kW}$ from latent heat transfer only. Modelling of the maximum possible cooling power of the condenser supports this measurement-based extrapolation, see appendix~\ref{appsubsec:ModellingTubeGeometry} for an estimation of the upper limit of the condenser heat transfer power. When operated at $\SI{1.5}{bara}$ LN$_{2}$ pressure, the total cooling power, based on the enthalpy change of the nitrogen, is $\SI{\sim 8}{kW}$ at a typical GN$_{2}$ flow of $\SI{1000}{slpm}$, assuming the GN$_{2}$ exits the system at $\SI{273.15}{\kelvin}$. It is demonstrated in section~\ref{subsec:CoolingpPerformancePressureStability} that GN$_{2}$ flow rates of this magnitude can be reached. However, since the test bed does not allow to stress the system with a heat load of $\SI{8}{kW}$, the maximum cooling power of the system can only be tested for a limited time in non-equilibrium conditions.

While left void for the subsequently described tests, the DS"~20k cryogenics system will include a radon trap using charcoal pellets along with a $\SI{0.003}{\micro\meter}$ bulk in-line gas filter (model SI2N0100B from Entegris~\cite{Entegris}) between HE~2 and HE~1 (see sections~\ref{sec:RequirementsConceptDesign} and~\ref{subsec:OngoingActivitites}). These components remove radon-isotopes and any residual particulates from the argon flow.

Eight pneumatically-actuated cryogenic valves~(model 18"~1603 from STÖHR~\cite{Stoehr}, DN~25 and PN~25) are installed to direct the flow path and isolate dedicated components. The two valves XV~A1 and XV~A5 through which the cryostat and the condenser can be directly connected are normally open~(NO), see figure~\ref{fig:UArCryoTestPID}. All other valves are normally closed~(NC). This ensures uninterrupted cooling provided to the cryostat and isolation of the GAr recirculation loop in case of a failure of the power or pneumatic system. As can be seen in figure~\ref{fig:UArCryoTestPID}, numerous resistance temperature detector~(RTD) probes (labelled TT) and pressure transducers (labelled PIT (model WUD"~20 from WIKA~\cite{WIKA}) or PT) are installed to monitor the system. Several bellows have been included in the design to prevent thermal stresses from arising. The condenser box features two free ports on the return inlet and one free port downstream of the condenser. While blanked in the test bed, in DS"~20k, these ports will be used for LAr filling from and recovery to storage containers (cf.~figure~\ref{fig:DS-20kConceptualPID}).

\subsection{Two-phase argon heat exchangers in cryostat}
\label{subsec:TwoPhaseHE}

\begin{figure}[t]
\centering
\begin{subfigure}[b]{0.48\textwidth}
\centering
\includegraphics*[height=0.24\textheight]{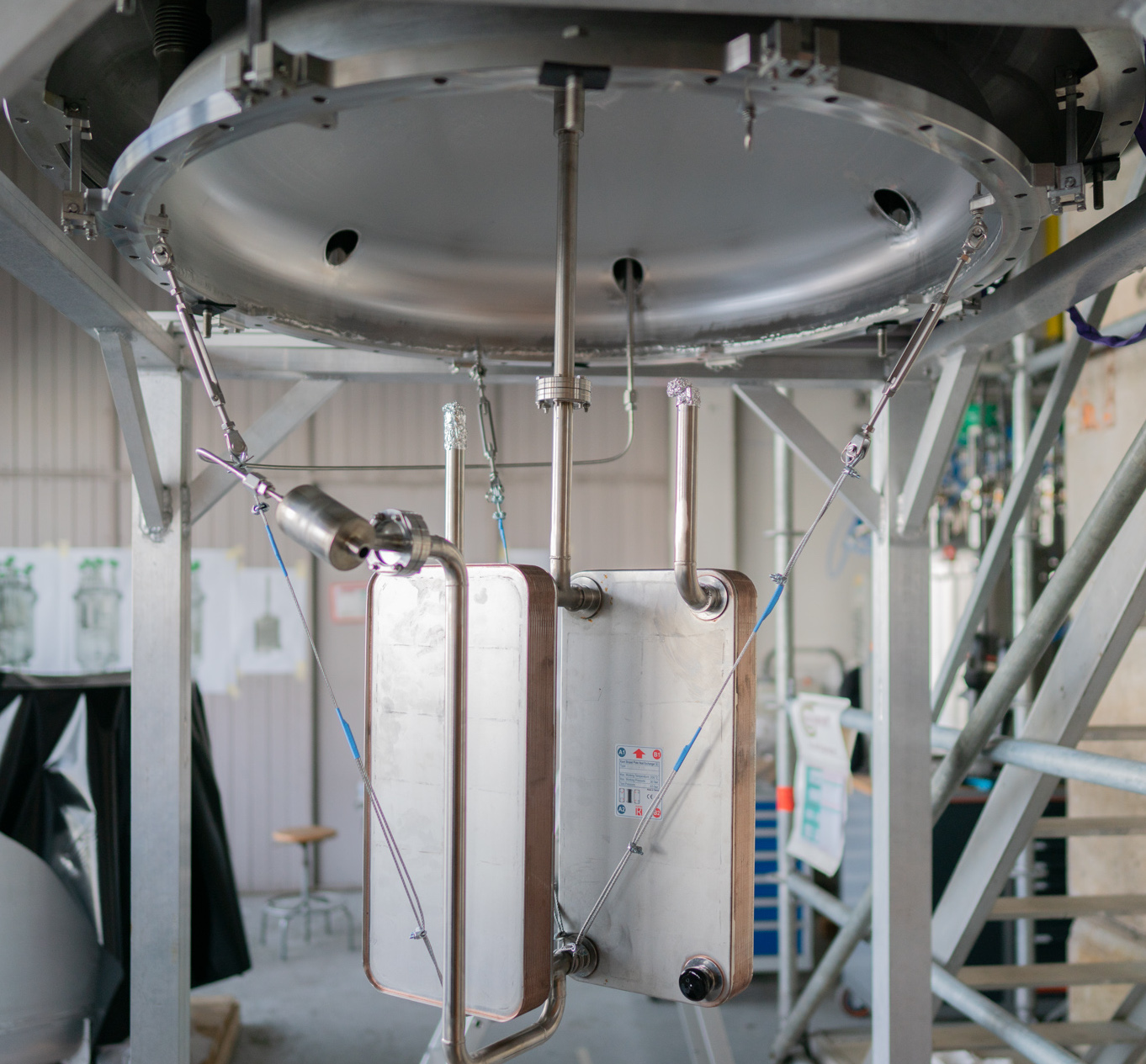}
\end{subfigure}
\quad
\begin{subfigure}[b]{0.48\textwidth}
\centering
\includegraphics*[height=0.24\textheight]{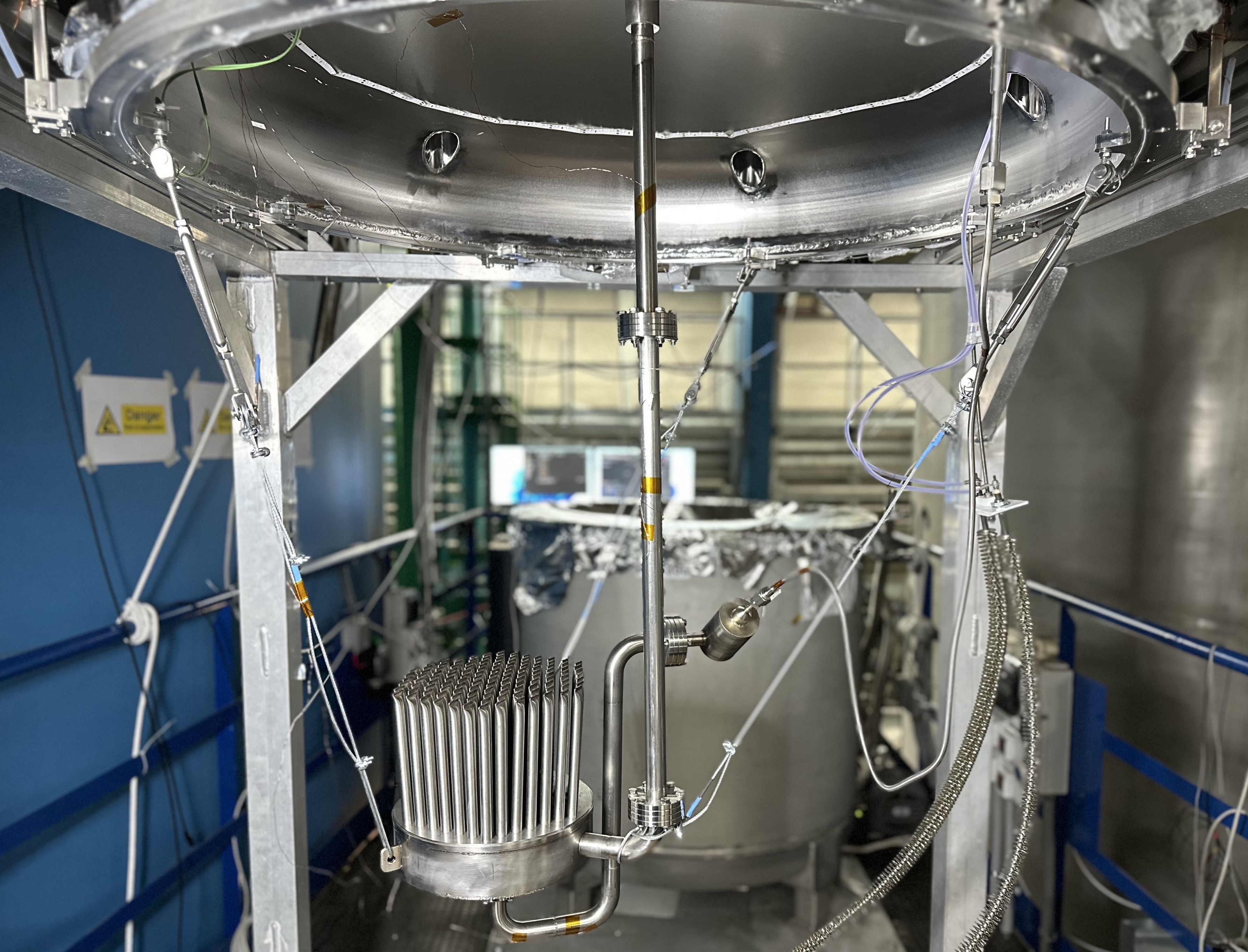}
\end{subfigure}
\caption[]{Pictures of the two-phase heat exchangers inside the cryostat. (Left): Parallel plate geometry. (Right): Multi-tube geometry.}
\label{fig:TwoPhaseHE}
\end{figure}

GAr for recirculation is boiled-off by heat which is transferred from the supplied higher-pressure argon. The heat transfer is driven by the temperature difference (and thus pressure difference) of the supply and the return side of the two-phase heat exchanger. One of the goals of the testing campaign is a measurement of the heat transfer coefficient, and thus of the required exchange area, and a comparison of the performance of heat exchanger geometries. We have tested two different configurations. The first consists of two plate heat exchangers (model K205-22 from KAORI~\cite{KAORI}) with 22~plates each (thickness: $\SI{0.6}{mm}$) and a total nominal heat exchange area of $\SI{4.84}{m^2}$, see figure~\ref{fig:TwoPhaseHE}~(left). The two heat exchangers are connected in parallel and used in counterflow configuration, i.e.~the flows are anti-parallel. The second configuration consists of a multi-tube geometry identical to the condenser from the preceding section with a nominal exchange area of $\SI{0.9}{m^2}$ (see figure~\ref{fig:TwoPhaseHE}~(right)). The two-phase argon heat transfer rate of this geometry was modelled and measured at small scale with 19~tubes, see appendix~\ref{app:MultiTubeGeometry}. Due to the use of non-insulated piping inside the cryostat and due to the heat exchanger shells, the surface available for heat transfer between the supply argon and the argon inside the cryostat is larger than the nominal exchange surface. Assuming that the heat transfer processes over the additional exchange surfaces $A_{i}$ with thicknesses $d_{\mathrm{Wall},\, i}$ are equal to the one over the nominal surface $A$ with thickness $d_{\mathrm{Wall}}$, we can define an effective exchange area as:
\begin{equation}
A_{\mathrm{eff}}\coloneq \left(\frac{A}{d_{\mathrm{Wall}}}+\sum_{i} \frac{A_{i}}{d_{\mathrm{Wall},\, i}}\right)\cdot d_{\mathrm{Wall}} \quad.
\end{equation}
The effective exchange areas of the two plate heat exchangers and the multi-tube heat exchanger are $\SI{4.94}{m^2}$ and $\SI{0.97}{m^2}$, respectively.

\subsection{Gas recirculation loop}
\label{subsec:RecirculationLoop}

\begin{figure}[t]
\centering
\includegraphics[width=0.618\textwidth]{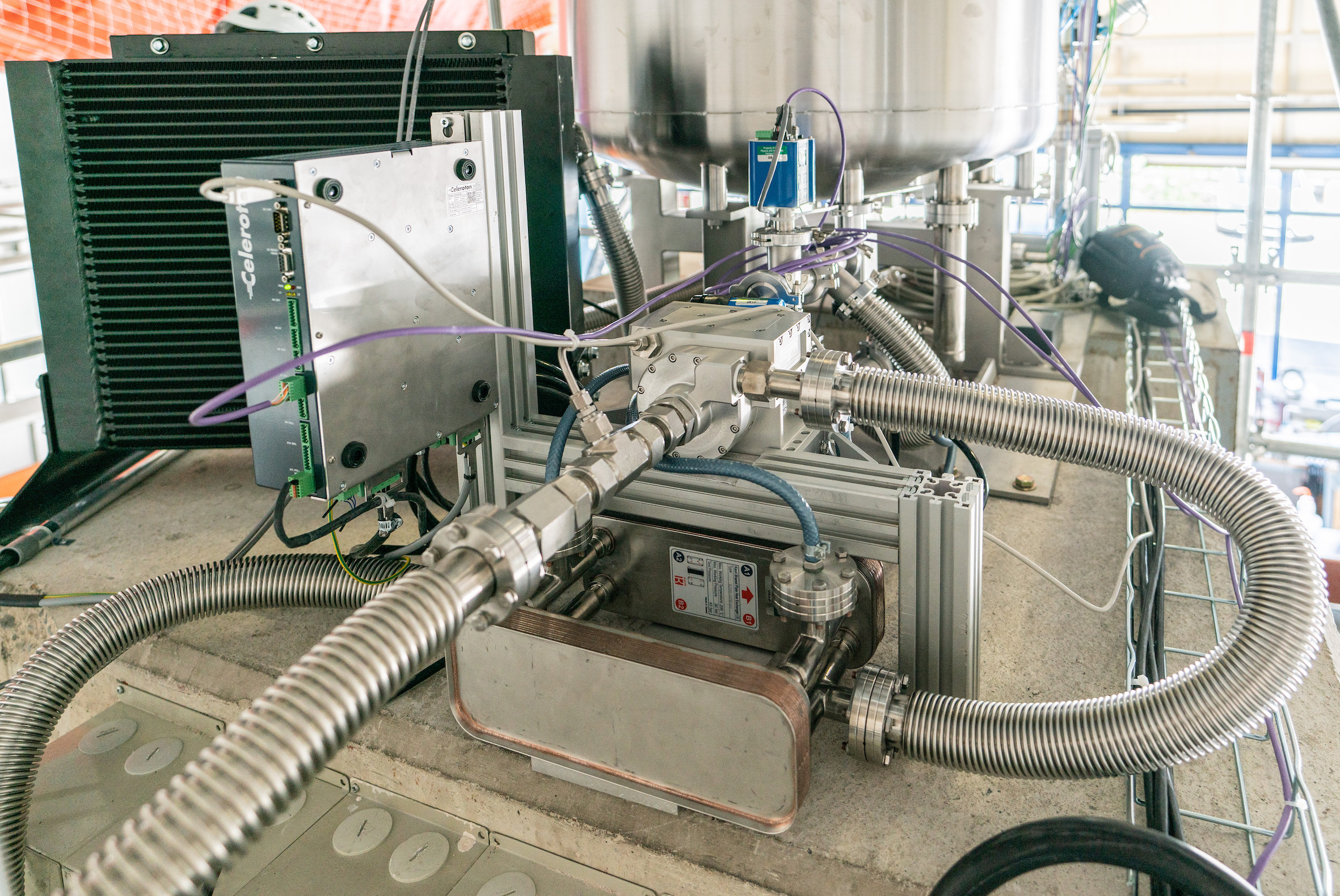}
\caption[]{Picture of the Celeroton compressor and the water/glycol-cooling as installed at CERN.}
\label{fig:CompressorRecirculationLoop}
\end{figure}

We refer to the warm GAr circuit outside of the condenser box as the gas recirculation loop, see figure~\ref{fig:UArCryoTestPID}. It features a radial-turbo compressor (model CT"~1000"~Ar) with a gas bearing from Celeroton~\cite{Celeroton}, see figure~\ref{fig:CompressorRecirculationLoop}, providing a peak pressure ratio of $1.6$ at an argon flow rate of $\SI{\sim 500}{slpm}$ at its maximum rotation frequency of $\SI{180}{krpm}$, according to manufacturer specifications (at $\SI{1.12}{bara}$ and $\SI{298.15}{\kelvin}$ inlet pressure and temperature). The GAr is chilled downstream of the compressor with water/glycol-cooled heat exchangers~(HE~W). The GAr flow rate is measured by an in-line mass flow meter (QuadraTherm\textsuperscript{\textregistered} 780i series from SIERRA Instruments~\cite{SIERRA}), which allows for argon flow measurements of up to $\SI{1500}{slpm}$. The $\SI{\sim 10}{m}$~circuit is made of $\SI{1}{in}$"~OD tubing and equipped with several high-purity pressure indicating transducers~(PIT, model WUD"~20 from WIKA~\cite{WIKA}).

\subsection{Cooling control}
\label{subsec:CoolingControl}

The cooling power is set by a closed-loop control based on the GAr pressure of the cryostat (process value) through the GN$_{2}$ outflow of the condenser (manipulated variable). As can be seen in figure~\ref{fig:UArCryoTestPID}, this can be realised with two independent flow control valves as control elements~(FV~PIDN and FV~PN) that are installed in parallel on the GN$_{2}$ vent line. The proportional valve FV~PIDN (model R18"~1603 from STÖHR~\cite{Stoehr}) is actuated pneumatically by an electro-pneumatic positioner (model SIPART~PS2 from SIEMENS~\cite{Siemens}) based on the current output of a PID~(proportional-integral-derivative) controller which receives the proportional voltage signal of the PIT~C1 on the cryostat. 

\begin{figure}[t]
\centering
\includegraphics*[height=0.618\textwidth]
{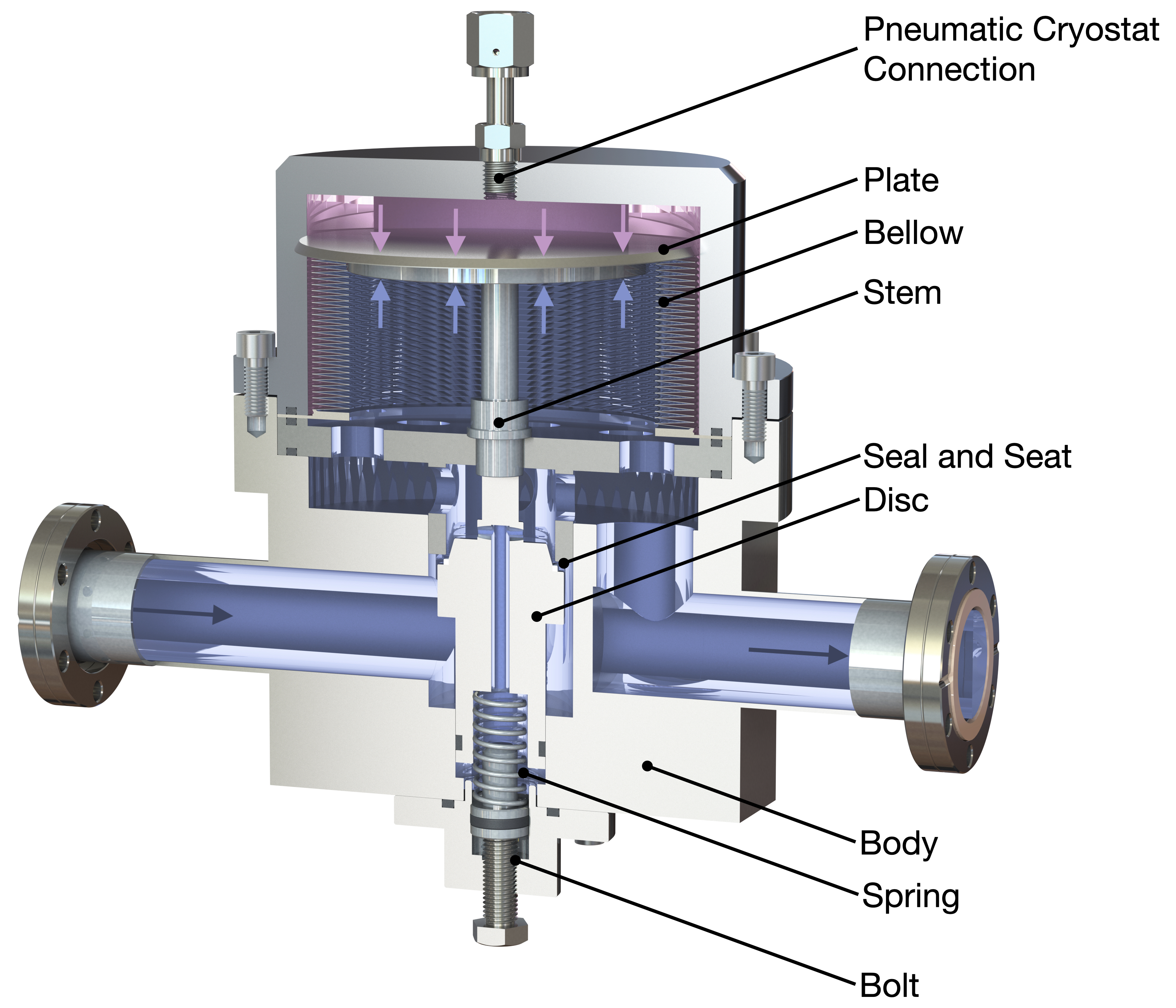}
\caption[]{Cross-section rendering of the mechanical safety control valve (\textit{bellow valve}). The purple region marks the GAr, the blue one the GN$_{2}$ volume.}
\label{fig:CoolingControlValve}
\end{figure}

The other flow control valve FV~PN, the \textit{bellow valve}, is actuated directly by the pneumatic feedback of the cryostat pressure. The pressure is applied onto a plate with a diameter of $\SI{\sim 130}{mm}$ which is suspended by an edge-welded stainless steel bellow, see figure~\ref{fig:CoolingControlValve}. The plate is mechanically coupled via a stem with the disc and the seal of the valve. If the cryostat pressure rises, the disc moves downward, opening the valve. A compression spring provides a counter force which pushes the stem back if the cryostat pressure decreases. A bolt at the bottom of the valve assembly can be used to adjust the pretension of the spring and thus, to calibrate the pressure setpoint of the cryostat. The GN$_{2}$ pressure downstream of the seat is applied on the other side of the bellow plate, balancing the GAr pressure. This direct feedback dampens the actuation of the valve and allows to operate the valve smoothly in a large dynamic range of GN$_{2}$ flows ($\SIrange{0}{1000}{slpm}$) with little absolute fluctuations of a few slpm. Even though the dimensions of the valve, with a seat diameter of $\SI{1.5}{in}$, are optimised for a high flow coefficient, it is possible to operate the valve precisely at flows of a few slpm and thus, at minimal system cooling power. This feedback feature makes the design similar to a pressure-reducing regulator. Unlike its installation in DS"~20k, in the test bed the valve is on its downstream side open to the atmosphere, and thus the GAr pressure setpoint is chosen relative to ambient pressure. The volume in which the spring moves is coupled by a channel to the downstream side of the valve, allowing GN$_{2}$ to enter and escape. This avoids an additional stem-position-dependent force due to a pressure differential of this volume and the internal bellow volume. The operation of the valve is widely insensitive to the upstream pressure. In section~\ref{subsec:Implications} we comment on upgrades of the bellow valve design for use with UAr in DS"~20k. 

During normal operations, the cooling control in DS"~20k will be handled via the PID-controlled valve FV~PIDN, which is normally closed. The valve XV~N just before the bellow valve FV~PN is normally open. Hence, in case of a failure of the power or of the pneumatic system supplying the pneumatic valves, the cooling control will fall back to the bellow valve FV~PN. This control valve is therefore intended as a passive safety backup, with regard to DS"~20k. It operates purely mechanically and requires no external supplies.

The other components on the cooling control line are, in the order of flow direction, an air heat exchanger with a fan as a heat sink to warm up the GN$_{2}$ exiting the condenser box to room temperature and a mass flow meter~\cite{SIERRA} downstream of the control valves, cf.~figure~\ref{fig:UArCryoTestPID}. In addition, there is a pressure indicating transducer~\cite{WIKA} before the heat sink. After the heat sink, there is a pressure relief valve and another pressure transducer installed. The heat sink is not strictly necessary during normal operations as the GN$_{2}$ is warmed up by heat exchange with the supply argon side. However, in case of a high cooling power demand at low or no GAr flow, the temperature of the exiting GN$_{2}$ could still be low and lead to the accumulation of moisture and ice around the non-insulated cooling control unit.

\subsection{Slow monitoring system}
\label{subsec:SlowMonitoring}

\begin{figure}[t]
\centering
\includegraphics[width=0.618\textwidth]{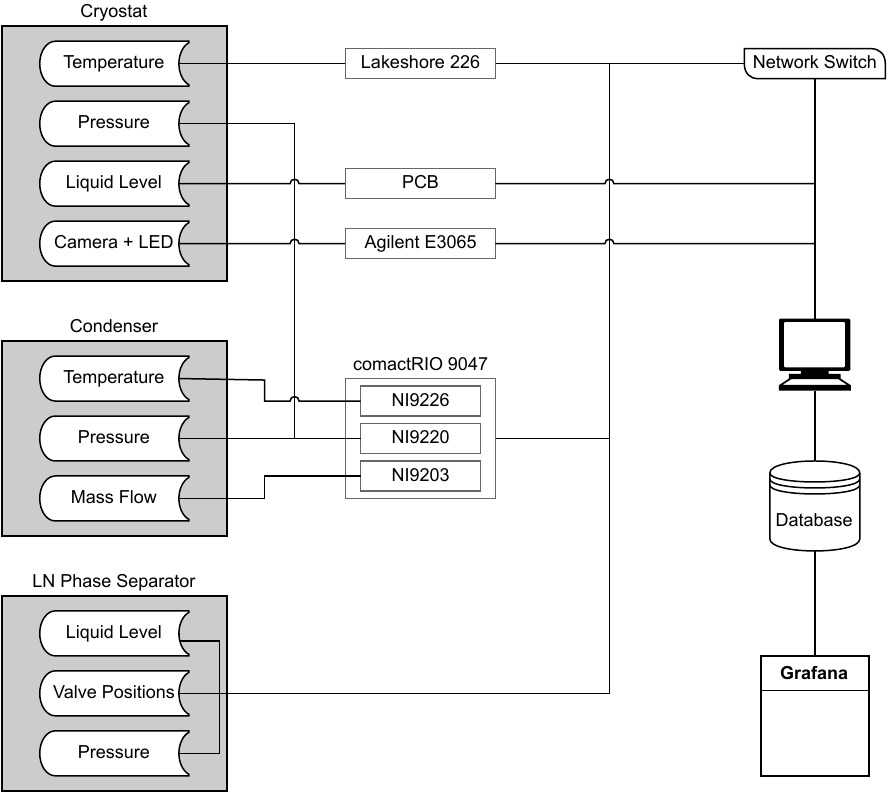}
\caption[]{Overview of the slow monitoring system for the cryogenic setup at LNGS.}
\label{fig:SlowMonitoring}
\end{figure}

The slow monitoring system that was implemented for this work passively and continuously monitors all major components of the system and provides a remotely accessible view of all sensor data, see figure~\ref{fig:SlowMonitoring}. The data acquisition is done via a National Instruments compactRIO~9074~\cite{compactrio9074}. A proprietary LabVIEW~\cite{labview} code saves the data to disk, which is then pushed to an InfluxDB~\cite{influxdb} database. The data monitoring and visualisation as well as the alerting of operators in case of anomalous sensor values are managed through a custom Grafana~\cite{grafana} interface. In addition to the slow monitoring, the system enables the operator to remotely adjust various operational parameters, which allows the running of the system with minimal on-site presence of operators. While this framework is deployed in the test bed, the DS"~20k slow monitoring and control system is discussed briefly in section~\ref{subsec:OngoingActivitites}.

\subsection{Testing facilities}
\label{subsec:Facilities}

Below, we present the testing facilities and infrastructure.

\paragraph{Cryolab at CERN}

\begin{figure}[t]
\centering
\begin{subfigure}[b]{0.6\textwidth}
\centering
\includegraphics*[height=0.3\textheight]{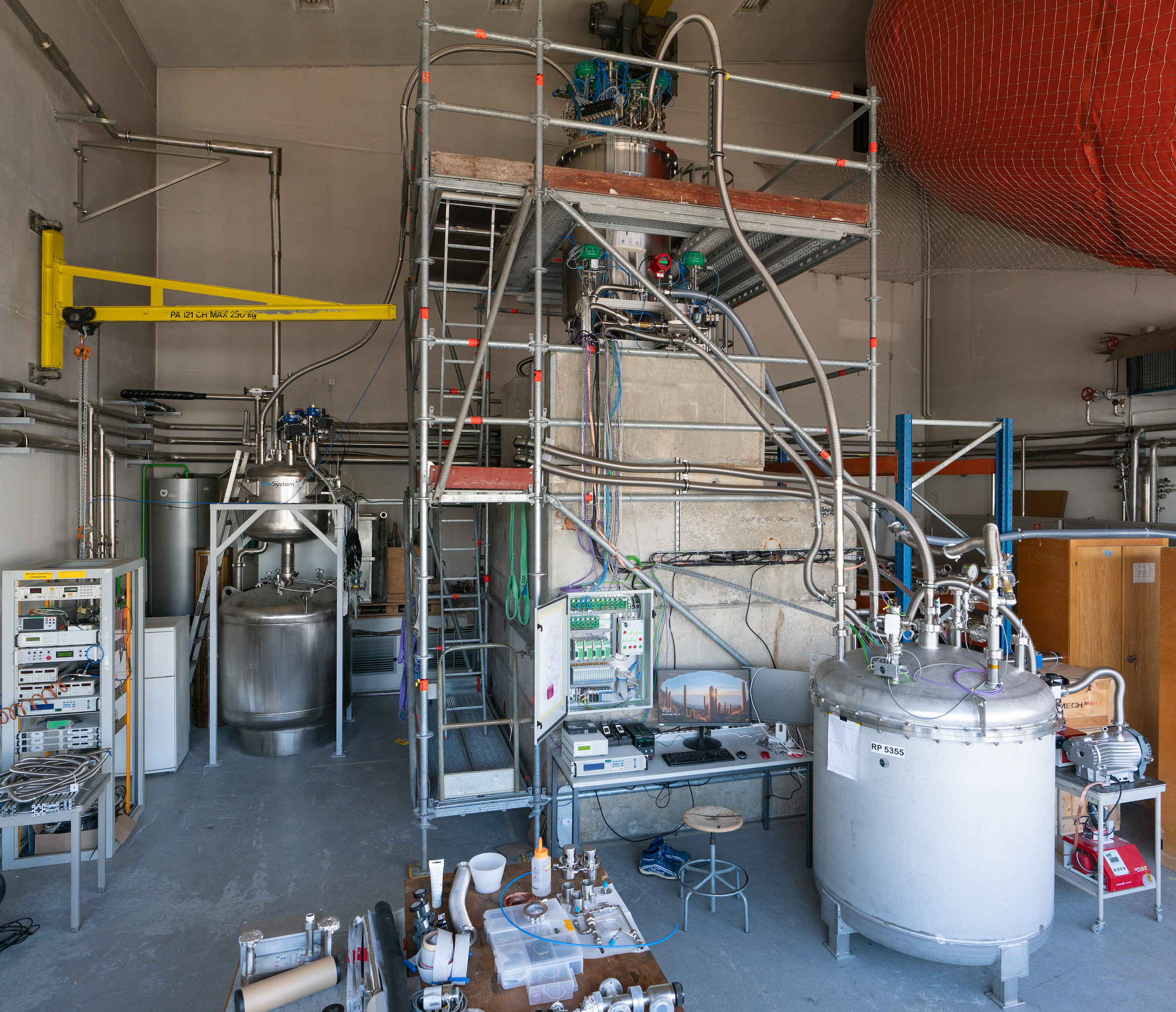}
\end{subfigure}
\quad
\begin{subfigure}[b]{0.36\textwidth}
\centering
\includegraphics*[height=0.3\textheight]{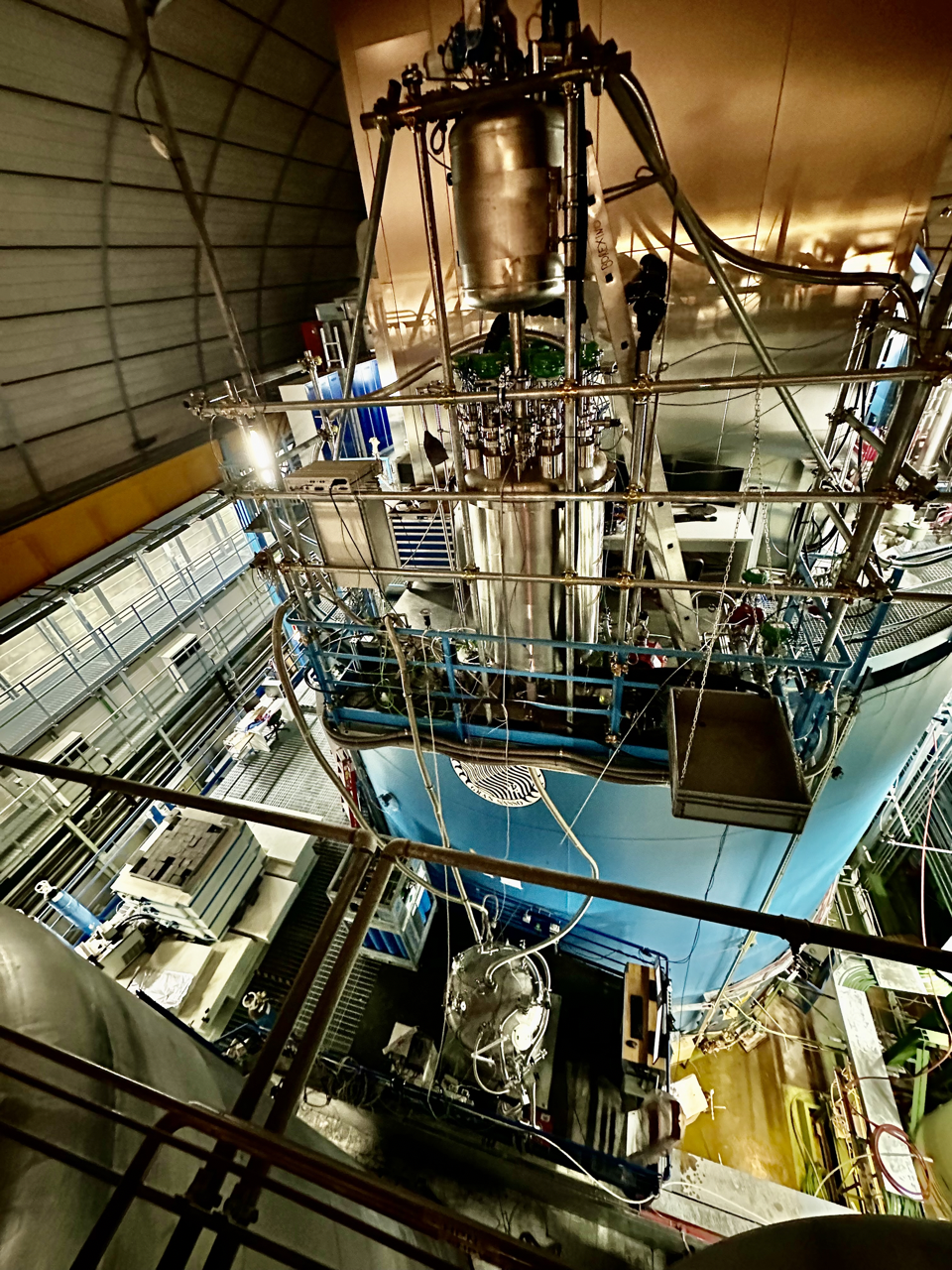}
\end{subfigure}
\caption[]{Testing facilities. (Left): CERN Cryolab. (Right): Hall"~C of the LNGS underground laboratories.}
\label{fig:Facilities}
\end{figure}

Between 2018 and 2022, the test bed and integral components of the DS"~20k UAr cryogenics system were fabricated, constructed, and commissioned at the CERN Cryolab. The commissioning focused on achieving pressure stability, optimising the cooling control, and performing first efficiency measurements and recirculation tests~\cite{Thorpe:2022dhw}. The test bed was served by a $\SI{660}{L}$ nitrogen phase separator. A picture of the facility in building~159 is shown in figure~\ref{fig:Facilities}~(left).

\paragraph{Hall-C at LNGS}

Following the initial testing effort at CERN, the system was shipped to LNGS. With the goal of rigorous benchmarking, it was set up underground in Hall~C on two platforms of the water tank of the Borexino Counting Test Facility~\cite{Alimonti:1998nt}, as shown in figure~\ref{fig:Facilities}~(right). A new nitrogen phase separator was positioned on top of the condenser box and operated at a nominal pressure of $\SI{1.5}{bara}$. It is supplied with LN$_2$ via an upstream insulated line of $\SI{\sim 100}{m}$ length. Operations started in mid October~2023 and lasted until the end of February~2024, consisting of two runs, one with each heat exchanger configuration.

\paragraph{DarkSide-20k Mockup TPC detector}

Before its final integration into the DS"~20k facility, the cryogenics platform at LNGS will be used to cool down a tonne-scale mockup of the DS"~20k TPC in late 2024. This scaled-down detector will share many design aspects with the DS"~20k TPC, as it is made of PMMA and features an octagonally shaped TPC with an inner diameter of $\SI{\sim 50}{\centi\meter}$ and a maximum drift length of $\SI{\sim 56}{\centi\meter}$. The insights gained during the mechanical assembly, cool-down and electrostatic tests of this mockup TPC will help validate various design choices. Most importantly, the cooling of the delicate acrylic TPC needs to be performed in a slow and controlled way. The procedure must ensure a uniform contraction of the parts, avoiding large temperature gradients that may lead to high stresses in the acrylic and interference of detector components that have tight clearances. The goal is to cool the detector at a rate of not more than $\SI{2}{\kelvin\per\hour}$ up to the LAr boiling point by circulating cold argon gas. The capabilities of the cryogenics platform in this regard have been tested and will be discussed in more detail in section~\ref{subsec:CoolingpPerformancePressureStability}. The design, testing and results of the DS"~20k Mockup TPC will be the subject of a stand-alone publication in the future.
\section{Commissioning and benchmarking}
\label{sec:CommissioningBenchmarking}

The performance data presented in this section was acquired with the test bed at LNGS. The findings are crucial to benchmark the component's capabilities but also to finalise the design for deployment in DS"~20k. We describe the measurements and the obtained results below and comment on their implications for DS"~20k in the next section. 

\subsection{Compressor performance and circuit resistance}
\label{subsec:CompressorPerformanceCircuitResistance}

Operating the system with GAr at ambient temperature ($\SI{286.5}{K}$) allows us to determine the compressor performance and to measure the flow-dependent pressure drop over the individual components of the recirculation loop. Note that the recirculation path can be shortened via the valve XV~A1 (cf.~figure~\ref{fig:UArCryoTestPID}), bypassing the cryostat. While the supply line downstream of the condenser CD~1 can be closed through XV~A5, the return line will still be open in this mode, using the cryostat volume as a buffer connected in parallel. This short circuit mode is useful to decouple the pressure drops over the heat exchanger cascade from the transfer lines and the two-phase heat exchanger and it provides a circuit with an overall lower flow resistance. We have operated the system in this mode at a cryostat pressure of $\SI{1.06}{bara}$ using compressor speeds in the range of $\SIrange{10}{180}{krpm}$ in steps of $\SI{10}{krpm}$. In the standard mode, which includes the cryostat in the recirculation path, we have acquired data at $\SI{1.06}{bara}$ and $\SI{1.20}{bara}$ cryostat pressure. In this mode, the helium-actuated valve PV~He at the cryostat inlet can be used to apply an additional impedance to the system, which is helpful to map the compressor behaviour at lower mass flows. Within the tested flow regime and the capabilities of the pump, the valve was utilised to add an average pressure drop in the range $\SIrange{0.1}{0.5}{mbar/slpm}$. 

\begin{figure}[t]
\centering
\includegraphics*[width=1\textwidth]{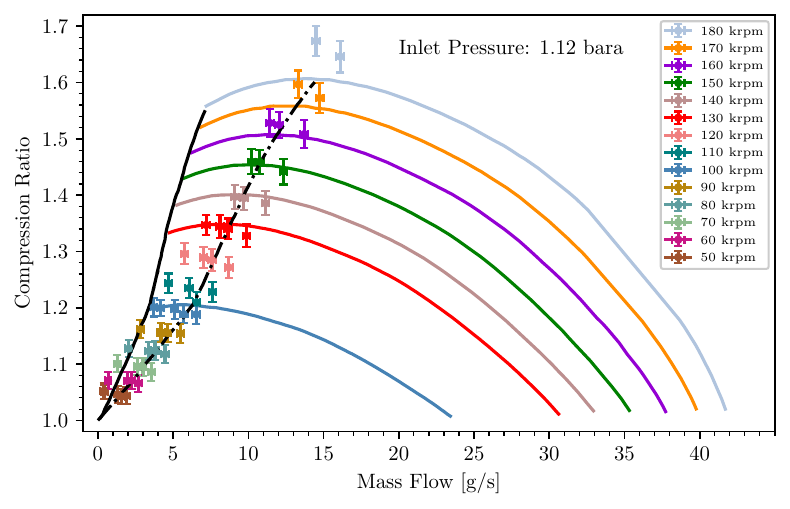}
\caption[]{Performance curve of the Celeroton CT"~1000"~Ar turbo compressor at $\SI{1.12}{bara}$ argon inlet pressure and various speeds. We omit to show data below $\SI{50}{krpm}$ rotational speed for the readability of the graph. The compression ratio at the maximum compressor speed peaks approximately at a flow rate of $\SI{500}{slpm}$ ($\SI{14.7}{g/s}$). The prediction from the manufacturer is shown as solid coloured lines. The solid and dash-dotted black lines indicate compressor surge and maximum compression, respectively.}
\label{fig:CompressorPerformance}
\end{figure}

In figure~\ref{fig:CompressorPerformance} we show the performance map of the turbo compressor, i.e.~the ratio of the outlet and inlet pressures versus the mass flow, for a compressor inlet pressure of $\SI{1.12}{bara}$. To extrapolate our measurements to this fixed inlet pressure, we took advantage of the fact that the compression ratio, for a given speed and recirculation circuit, is constant as a function of inlet pressure. The mass flow was extrapolated linearly to the mentioned inlet pressure utilising the data acquired at two different system pressures. We observe a good agreement with the manufacturer's prediction around the maximum compression line. The finite circuit resistance of the test bed, even if operated in the short circuit mode, restricts possible measurements to the highest mass flows shown in figure~\ref{fig:CompressorPerformance}. When extending the measurements towards the surge line, we have observed that above $\SI{100}{krpm}$ the pump's gas bearing became unstable before reaching the predicted end of the dynamic range. This can be observed by means of a metallic rubbing noise and eventually by an error of the pump controller. In our test, the leftmost points mark the limit where compressor surge starts to occur.   

\begin{figure}[t]
    \centering
    \includegraphics[width=0.618\textwidth]{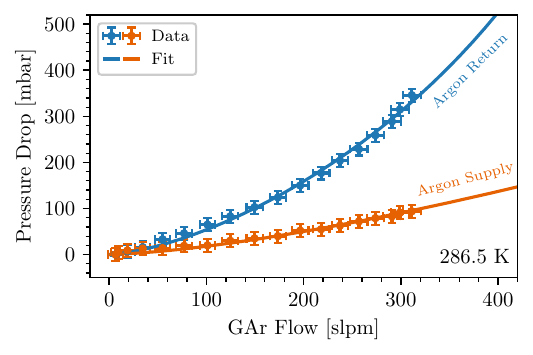}
    \caption[]{Measured pressure drop over the heat exchanger cascade in the centre circuit (supply) and an outer circuit (return) with argon gas at $\SI{286.5}{K}$ in short circuit mode, together with the model described in the text. The pressure on the argon side of the condenser was kept stable at $(1065 \pm 3)\,\si{mbara}$, which is possible due to its coupling to the cryostat buffer. This is the inlet (outlet) pressure of the argon return (supply) side.}
    \label{fig:CircuitResistance}
\end{figure}

The short circuit mode also allows us to measure the pressure drop over the heat exchanger cascade for the supply and return path with the pressure probes PIT~A1, A2, and CDA (cf.~figure~\ref{fig:UArCryoTestPID}). We obtain the result shown in figure~\ref{fig:CircuitResistance} and observe a higher pressure drop for the return path. This is due to the dual-circuit geometry of the heat exchangers described in section~\ref{subsec:CondenserBox}, which features more than twice as many channels in the centre circuit than in the individual outer circuit paths. The channel pressure drop can be modelled with the Darcy–Weisbach equation~\cite{Neagu2016}: 
\begin{equation}
\Delta p_\mathrm{ch}=4f\frac{\rho u^2}{2}\frac{L}{D_\mathrm{h}} \quad.
\end{equation}
Here, $f$ is the friction factor, $\rho$ is the mass density of the fluid, $u$ is the inter-plate fluid velocity, $L$ is the plate length relevant for the pressure drop, i.e.~the distance of the heat exchanger ports, and $D_\mathrm{h}$ is the hydraulic diameter. Abstracting the cross-section of the channels by a rectangle and accounting for the plate corrugation with a free parameter on the inner-plate distance, the hydraulic diameter and the mean fluid velocity can be defined in terms of the known heat exchanger geometry. We do not need to include a term for the difference between the bulk and wall viscosity here, as we are considering an isothermal process. To describe the chevron-type plate heat exchanger, we use the Muley friction factor~\cite{Muley1999}:
\begin{equation}
f=\left( \frac{\alpha}{30} \right)^{0.83} \left[ \left( \frac{30.2}{\mathrm{Re}} \right)^5 + \left( \frac{6.28}{\mathrm{Re}^{0.5}} \right)^5 \right]^{0.2} \quad,
\end{equation}
where $\alpha=\SI{60}{\degree}$ is the corrugation inclination angle relative to vertical direction and $\mathrm{Re}$ is the Reynolds number. In figure~\ref{fig:CircuitResistance}, we show the good agreement of this single-parameter model to the data. From the fit, we find that the effective inner-plate distance, as relevant for the flow, is reduced to $\SI{58}{\%}$ compared to the nominal distance that results from the plate size and the channel volume provided by the manufacturer. This is due to the corrugation of the plates. We have modelled the pressure drop for every heat exchanger stage in terms of $u$ and $\mathrm{Re}$ using the GAr density at the respective pressure conditions. Note that this test has been performed at $\SI{286.5}{K}$. Under operating conditions, the mass density of the gas is higher compared to the data presented here, and thus the pressure drop is expected to be generally lower at the same mass flow since $u \propto \rho^{-1}$ and $\mathrm{Re} \propto \rho $. In particular, the pressure drop will vary for the different heat exchanger stages, increasing from HE~1 to HE~5. 

The pressure drop over the GAr recirculation loop in the test bed is found to be dominated by the GAr-GAr heat exchanger cascade. Pressure drops in the warm tubing are not higher than $\SI{20}{mbard}$ at the tested flow rates. Note that the pressure drop over the coolant-chilled heat exchangers HE~W downstream of the compressor cannot be measured in lack of an intermediate pressure transducer. Thus, their flow resistance cannot be disentangled from the measured compression ratio of the compressor. 

Extending the flow test to the standard recirculation path, i.e.~including the cryostat, we can measure the pressure drop over the two-phase plate heat exchangers. We measure a maximum pressure drop of $\SI{70}{mbard}$ at $\SI{288}{slpm}$ and $(1065 \pm 3)\,\si{mbara}$ cryostat pressure between the condenser and the cryostat when the helium-operated pneumatic valve PV~He is fully opened. Assuming a negligible contribution of the transfer line between the condenser box and the test cryostat, and of the valve PV~He, this pressure drop is entirely due to the two-phase plate heat exchangers CD~2. The flow-dependence of the pressure drop in the two-phase plate heat exchanger is similar to the one shown for the gas heat exchanger cascade in figure~\ref{fig:CircuitResistance}.

\subsection{Cooling performance and pressure stability}
\label{subsec:CoolingpPerformancePressureStability}

The benchmarking runs of the system were conducted with a nominal GN$_{2}$ pressure of $\SI{1.5}{bara}$, measured in the phase separator with the pressure transducer~PT~PS. The nitrogen saturation temperature at this pressure is $\SI{80.8}{K}$~\cite{NIST}. Depending on the liquid level setpoint of the phase separator, the nitrogen pressure in the condenser is slightly higher at the order of $\SI{10}{mbard}$. Even though the system is designed to operate with an unpressurised nitrogen supply, and hence to work with an LN$_{2}$ temperature below the argon's triple point, a certain overpressure with respect to the nitrogen vent downstream of the cooling control is required in order to allow for high GN$_{2}$ flows, i.e.~for high cooling power. To demonstrate the system's robustness against LN$_{2}$ supply pressure changes, we have operated the cooling control at various other phase separator pressures in the range $\SIrange{1.2}{2.0}{bara}$. It can be calculated that the associated changes of the LN$_{2}$ enthalpy give rise to percent-level changes in the GN$_{2}$ flow, which remain unnoticed at the cooling power required in normal operation mode and associated GN$_{2}$ flow rates of the order of $\SI{10}{slpm}$. The cooling control with both the bellow valve~FV~PN and the PID-controlled valve~FV~PIDN remained stable and the cryostat pressure remained unchanged at all tested pressures. The temperature of the LAr dripping down the condenser was above the triple point at all times and was found to be relatively insensitive to changes of the nitrogen pressure. For the tested range of nitrogen pressures, we observed only small variations of the order of $\SI{0.1}{K}$ around its nominal value of $\SI{84.4}{K}$, which is present at $\SI{1.5}{bara}$ nitrogen pressure.

\begin{figure}
    \centering
    \includegraphics*[width=0.8\textwidth]{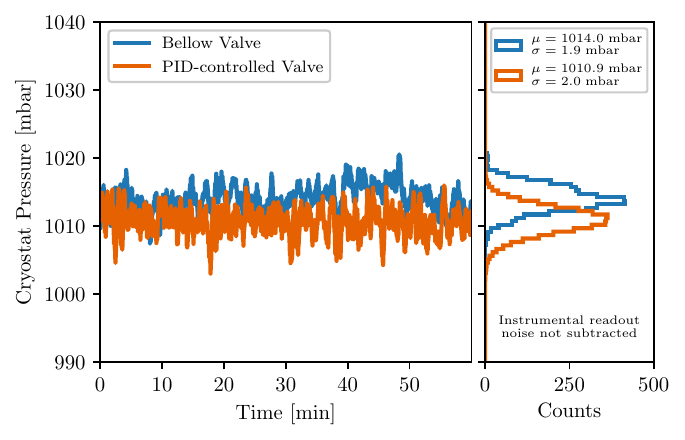}
    \caption[]{Argon pressure inside the cryostat as a function of time. The instrumental readout noise dominates the overall pressure fluctuations and is not subtracted here. See the text for more details.} 
    \label{fig:PressureStability}
\end{figure}

As outlined in section~\ref{sec:RequirementsConceptDesign}, the pressure stability of the system is generally of great importance for a stable operation of the gas pocket, and thus for the uniformity of the S2 gain as a function of time. However, it should be noted that pressure fluctuations of the gas ullage will not lead to liquid level changes due to the incompressibility of LAr nor to changes in the gas pocket thickness. As mentioned, the gas pocket is created either in-situ by boiling liquid or by a direct gas feed from the purification system. The pressures of both gas sources follow changes of the ullage or overall system pressure. The total pressures in the gas pocket and in the liquid directly below are therefore equal and change synchronously. The maximum gas pocket thickness is defined through a gas overflow over the edge of an upside-down bucket. The thickness is therefore constant at its maximum as long as enough gas is supplied. Small changes in the gas pocket pressure do also not alter the S2 signal gain significantly due to an approximate linear relationship of the reduced (by atom number density) electroluminescence yield and the reduced electric field~\cite{Monteiro:2008zz,Oliveira:2011xx,Zhu:2018tqu}. While the pressure stability is a system-dependent quantity and cannot directly be translated from the test bed to DS"~20k, it is a good indication of the level of stability that is possible.

Figure~\ref{fig:PressureStability} shows the pressure stability of the system over a $\SI{60}{min}$ window when using either the bellow valve or the PID-controlled valve for the cooling control during stable operations. The pressure inside the cryostat in both cases can be kept stable to within $\SI{\sim 2}{\milli\bar}$ RMS reading. A low-pass filter has been applied to the data to remove high-frequency readout noise. The observed pressure fluctuations are largely dominated by instrumental readout noise which is to be expected based on the specifications of the pressure sensor in use with a dynamic range of $\SIrange{0}{7}{bara}$ and an accuracy of better than $\SI{10.5}{mbar}$ root sum square~\cite{WIKA}. A principal component analysis of the two datasets presented in figure~\ref{fig:PressureStability} in conjunction with two separate noise-only datasets (with $\sigma= \SI{1.8}{\milli\bar}$) yields a residual system-intrinsic pressure fluctuation for the bellow and PID-controlled valve of $\sigma \sim \SI{0.1}{\milli\bar}$ and $\sigma \sim \SI{0.2}{\milli\bar}$, respectively. The noise-only dataset is defined as the closest dataset to each run where the cryostat was under high vacuum and thus by definition contains only instrumental readout noise.

Pressure fluctuations of the phase separator sitting directly on top of the condenser box result in small spills through the chicken feeder. This can provide an unwanted additional cooling power at zero cooling power demand, especially during filling periods of the phase separator. This test was thus performed in between two refills to mitigate this effect. It is expected that, with DS"~20k, this side effect of the system will be significantly reduced due to the use of two nitrogen phase separators, which alternate between filling and cooling to maintain a stable nitrogen supply pressure (cf.~figure~\ref{fig:DS-20kConceptualPID}). These will be physically separated from the condenser box and connected via vacuum-insulated transfer lines which provide a larger buffer volume that is better in absorbing pressure fluctuations. It is worth noting that this effect has not been observed in the test bed at CERN in which the phase separator was located at a lower height than the condenser box. In this case, there is no liquid column maintained above the chicken feeder at zero cooling power demand. See section~\ref{subsec:Implications} for more details.

\begin{figure}
    \centering
    \includegraphics*[width=0.8\textwidth]{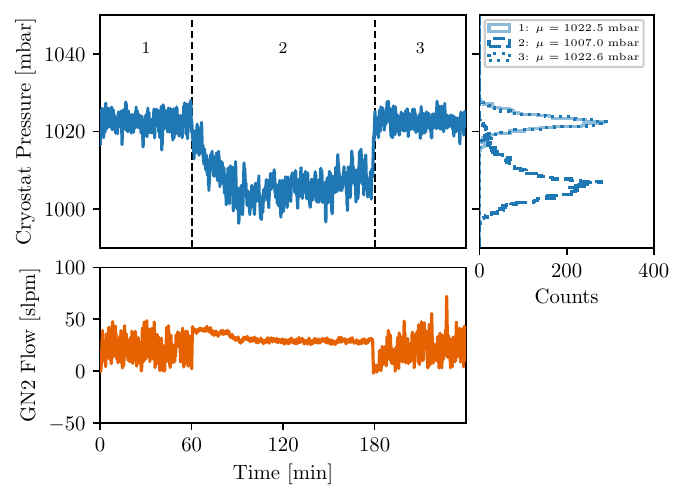}
    \caption[]{Cryostat pressure and GN$_{2}$ flow during switching of the control valves: 1 -- PID-controlled valve (FV~PIDN), 2 -- Bellow valve (FV~PN), 3 -- PID-controlled valve (FV~PIDN).}
    \label{fig:SwitchingBehaviour}
\end{figure}
\begin{figure}
    \centering
    \includegraphics*[width=0.8\textwidth]{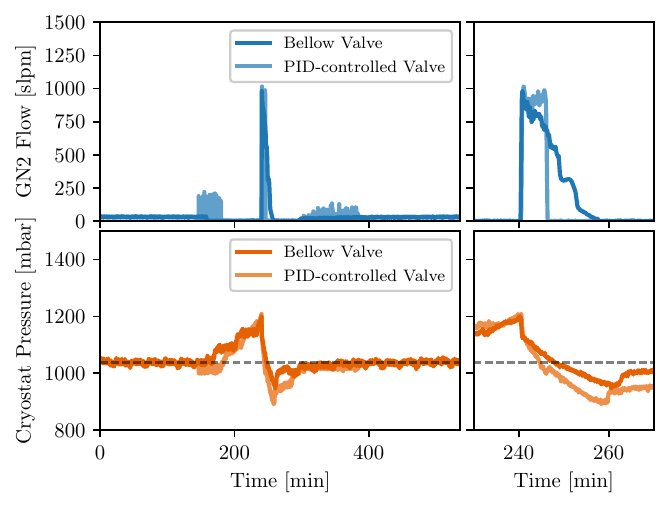}
    \caption[]{Response of the cooling control valves to an induced pressure increase. The dashed line denotes the mean nominal cryostat pressure before the pressure increases.}
    \label{fig:StepResponse}
\end{figure}
\begin{figure}[t]
    \centering
    \includegraphics*[width=0.8\textwidth]{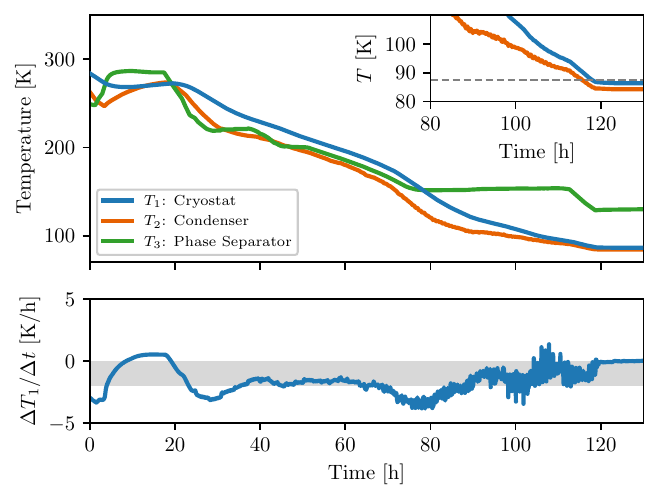}
    \caption[]{Temperature of the LAr in the cryostat, of the LAr downstream of the condenser, and of the GN$_2$ vent of the phase separator as a function of time during the cool-down of the system by recirculating cold argon gas. The insert in the top panel shows the moment LAr accumulates in the cryostat, detected by the RTD which is placed just above the cryostat floor. The grey region in the bottom panel shows the target cool-down rate of $\SI{-2}{K/h}$.}
    \label{fig:CoolingRate}
\end{figure}

In figure~\ref{fig:SwitchingBehaviour} the response of the system to the switching from the PID-controlled valve to the bellow valve and back is demonstrated. Two features are worth pointing out: firstly, the setpoints of the valves were not calibrated and aligned, which resulted in a pressure change after the switching. However, the transition is smooth and no large pressure spikes are introduced into the system. Secondly, the N$_{2}$ flow, directly proportional to the cooling power, is significantly more stable during the operation with the bellow valve. This is related to the flow coefficient of the PID-controlled valve being oversized for the needs of the system at low flows, which allows for large cooling power but little dynamic range for small cooling power requirements. Therefore, more responsive PID parameters had to be chosen to stabilise the pressure. A more appropriately sized valve orifice would mitigate this behavior entirely as will be discussed in section~\ref{sec:DS-20kOutlook}. This will further improve the system's pressure stability.

Furthermore, both cooling control valves were subjected to an induced cryostat pressure increase of $\SI{\sim 200}{mbard}$ to test their ability to return the system to equilibrium, as shown in figure~\ref{fig:StepResponse}. The pressure inside the cryostat was increased by closing the GN$_2$ vent line of the cooling control, i.e.~by closing XV~N and FV~PIDN in figure~\ref{fig:UArCryoTestPID}. Upon reinstating the cooling control valves, the GN$_{2}$ flow rate increased to $\SI{\sim 1000}{\slpm}$, providing immediate and large cooling power to the system and reducing the pressure inside the cryostat to or below the equilibrium setpoint within $\SI{\sim 10}{\min}$. Within approximately $\SI{40}{\min}$ the system went back to equilibrium pressure. It should be noted that the UAr cryogenics system in DS"~20k is not expected to have pressure fluctuations on the order of $\SI{100}{mbard}$ in absence of system failures or catastrophic events, and when standard operating procedures are followed, but this test lends confidence to the system to reliably respond to smaller possible fluctuations during the normal operations.

The final test for the cooling control is related to the cool-down rate, which is only relevant in the case of a detector present inside the cryostat such as the DS"~20k TPC Mockup that was discussed in section~\ref{subsec:Facilities}. During the initial cool-down of the detector it has to be ensured that a pre-defined temperature gradient, in time and space, is not exceeded to keep the material stresses within the limits. In figure~\ref{fig:CoolingRate}, we show the cool-down of the test cryostat. The test was conducted using cold GN$_2$ in the phase separator and condenser only, thus avoiding condensation of argon. Only at the end of the test, the phase separator was filled with LN$_2$, leading to a slow accumulation of LAr at the bottom of the cryostat. The test was conducted with the test cryostat, containing only the two-phase heat exchanger, and that the RTD used to measure the temperature was freely hanging in gas. It is expected that a smoother gradient with a smaller magnitude can be maintained inside TPC materials with significant heat capacity.

\begin{figure}
    \centering
    \includegraphics*[width=1\textwidth]{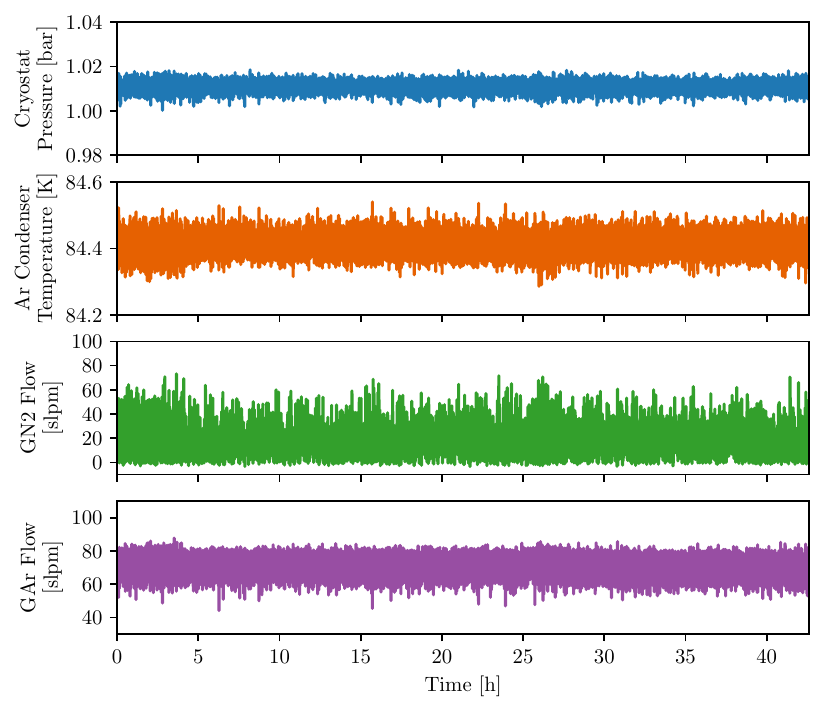}
    \caption[]{Stability of the system's key parameters during a $\SI{43}{h}$ time period using the PID-controlled valve.} 
    \label{fig:LongtermStability}
\end{figure}

Figure~\ref{fig:LongtermStability} shows the stability of the system's key parameters over $\SI{43}{h}$ using the PID-controlled valve for cooling control. Unlike the pressure stability test shown in figure~\ref{fig:PressureStability} that was performed in between two refills of the phase separator with a continuously decreasing LN$_2$ level, this window involves regular refills every $\SI{\sim 17}{min}$ to keep the LN$_2$ level at $45\substack{+1 \\ -2} \,\si{\%}$.

\subsection{System efficiency}
\label{subsec:SystemEfficiency}

In this section we will evaluate the dynamic heat load of the system, i.e.~the energy loss during recirculation. To assess the heat recovery capability of the system during recirculation, we define an efficiency:
\begin{equation}
\eta \coloneq 1-\frac{\rho_{\mathrm{N_{2}}}}{ \rho_{\mathrm{Ar}}} \left|\frac{\Delta h_{\mathrm{N_{2}}}}{\Delta h_{\mathrm{Ar}}}\right| \frac{\diff \mathcal{F}_{\mathrm{N_{2}}}}{\diff \mathcal{F}_{\mathrm{Ar}}} \quad.
\label{eq:eta}
\end{equation}
For both fluids, we denote the mass density at standard conditions ($\SI{273.15}{K}$, $\SI{1}{bara}$) by $\rho$, the specific enthalpy change by $\Delta h$ and the volumetric flow rate by $\mathcal{F}$. As described in section~\ref{subsec:CondenserBox}, heat transfer from argon to nitrogen takes place in both the condenser and the heat exchanger cascade. While the latent heat exchange, i.e.~the phase change happens in the former, gas-gas exchange happens in the latter. That is to say that the relevant enthalpy change for nitrogen ranges from liquid state to gas state at almost ambient temperature ($\SI{286.5}{K}$) and vice versa for argon. For this calculation we use the fluid properties at a typical nitrogen pressure of $\SI{1.5}{bara}$, as present in the condenser and an argon pressure of $\SI{1.06}{bara}$ as present in the cryostat~\cite{NIST}. It is clear that the pressures vary throughout the system. Those uncertainties alter the result however only at the per mille level.

With this definition for $\eta$, a system achieving $\SI{100}{\%}$ dynamic efficiency corresponds to an ideal configuration in which the nitrogen flow remains constant with the argon flow, compensating only the static heat load onto the system. In this scenario, argon is then recirculated with zero loss, i.e.~supplied argon is cooled and liquefied solely by the returning argon in the gas-gas and the two-phase heat exchangers. Conversely, $\SI{0}{\%}$ efficiency corresponds to a system in which the absolute values of the enthalpy changes of argon and nitrogen are equal.

\begin{figure}[t]
\centering
\includegraphics[width=0.618\textwidth]{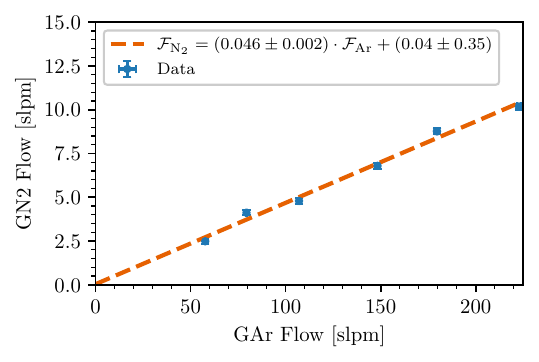}
\caption[]{N$_{2}$ flow vs.~Ar flow with linear fit. The slope of the curve is a measure of the efficiency of the system, i.e.~the dynamic heat loads. The static heat load of the system was measured by means of the initial argon boil-off rate during the cryostat venting at the end of operations and subtracted in this plot. As expected, the offset of the fit on the ordinate is thus compatible with zero.}
\label{fig:Efficiency}
\end{figure}

The efficiency of the system can be evaluated by measuring the increment in nitrogen flow in response to an increase in argon flow, attributed to the introduction of heat into the system. Figure~\ref{fig:Efficiency} shows the nitrogen flow as a function of the argon flow, fitted with a linear regression. This reduces $\diff \mathcal{F}_{\mathrm{N_{2}}} / \diff \mathcal{F}_{\mathrm{Ar}}$ to a constant and we obtain an efficiency of $\eta = \SI{95}{\%}$.

It is important to note that the data presented here included argon phase change, verified visually at the supply outlet into the cryostat~(PV~He) and by means of flow and pressure data. It is clear that gas can be recirculated at high speed with much lower dynamic loss (cf.~preliminary tests in reference~\cite{Thorpe:2022dhw}). To ensure that the system was in thermal equilibrium for the measurements, visible as a convergence of the temperatures, we allowed the system to relax for $\SI{\sim 12}{h}$ after flow changes. The efficiency measurement was performed with the bellow valve FV~PN for the GN$_2$ flow control. 

\subsection{Maximum argon flow rate and two-phase heat exchanger sizing}
\label{subsec:MaximumFlowTwoPhaseHESizing}

\begin{figure}[t]
\centering
\begin{subfigure}[b]{0.47\textwidth}
\centering
\includegraphics*[width=1\textwidth]{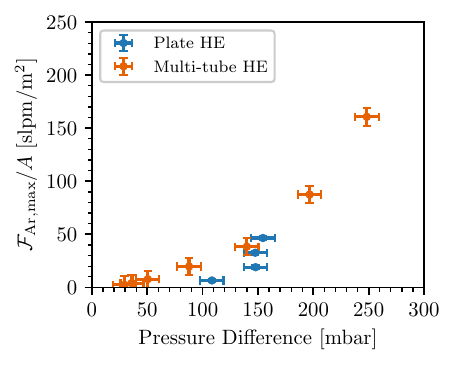}
\end{subfigure}
\begin{subfigure}[b]{0.51\textwidth}
\centering
\includegraphics*[width=1\textwidth]{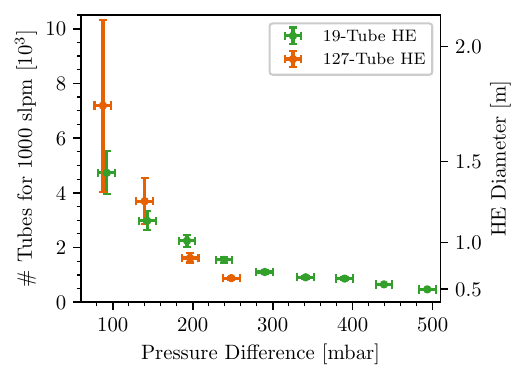}
\end{subfigure}
\caption[]{(Left): Maximum achieved argon flow~$\mathcal{F}_\mathrm{Ar,max}$, normalised by exchange surface~$A$, vs.~pressure difference for both heat exchanger geometries. The argon flow caused by static heat load has been subtracted. (Right): Number of $\nicefrac{1}{2}\,\si{in}\times \SI{7}{in}$ tubes required for $\SI{1000}{slpm}$ argon flow and heat exchanger~(HE) diameter vs.~differential pressure. We show the data from the extrapolation based on a small-scale test with a 19"~tube heat exchanger (see appendix~\ref{appsubsec:MultiTubeSmallScaleTest}) and based on data acquired with the 127"~tube heat exchanger. The results obtained with the 19"~tube heat exchanger were scaled to a tube length of $\SI{7}{in}$.}
\label{fig:MaxArFlowNbTubesDS20k}
\end{figure}

Both two-phase heat exchanger configurations have been tested with regard to their heat transfer capabilities. The heat transfer, and hence the condensation rate increases with the pressure difference between the two sides of the heat exchanger. The helium-actuated pneumatic valve PV~He downstream of the two-phase heat exchanger was used to create an alterable pressure difference in order to map this relation. In figure~\ref{fig:MaxArFlowNbTubesDS20k}~(left) we show the results for the two configurations, normalised per exchange surface to ease comparison\footnote{Note that a static heat load on the cryostat, compensated by cooling in the condenser, leads to a positive offset flow which has been subtracted from the data. The static cryostat heat load was determined by three independent measurements with consistent results: from the ordinate axis intercept of the nitrogen versus argon flow (see section~\ref{subsec:SystemEfficiency}), from the convective argon flow when the compressor is off and from the initial argon vent flow when the system was shut down.}. The error bars include both the instrumentation accuracy and standard uncertainties on the pressure and flow means. While for the plate heat exchangers, the compressor performance limited the measurement to go just slightly above $\SI{150}{mbard}$ pressure difference, the heat transfer capability of the multi-tube heat exchanger could be mapped up to almost $\SI{250}{mbard}$. While below $\SI{150}{mbard}$, the plate heat exchangers provide a lower maximum argon flow, there is a sharp increase visible towards the high end of the measured range of pressure differences. The lower performance might be attributed to the small inter-plate distance of the plate heat exchangers which are not designed for phase change processes. The effectively available exchange area can get reduced due to the displacement of LAr by boiled-off GAr. For the multi-tube heat exchanger instead, boiled-off gas can quickly rise due to the large tube pitch which is expected to enhance the heat transfer coefficient through induced forced convection. In order to determine the size of the multi-tube heat exchanger required to reach the DS"~20k target of $\SI{1000}{slpm}$, we extrapolate based on the results obtained with a 19"~tube (see appendix~\ref{appsubsec:MultiTubeSmallScaleTest}) and the 127"~tube heat exchanger. The corresponding plot is shown in figure~\ref{fig:MaxArFlowNbTubesDS20k}~(right). Here, we omit to include data from the 127"~tube heat exchanger below $\SI{50}{mbard}$ due to the large relative uncertainties associated with the flow meter accuracy at low flows. Overall, we observe a compatible result below $\SI{150}{mbard}$. Above this value, the extrapolation based on the results with the 127"~tube configuration yields a smaller required heat exchanger for DS"~20k. This implies that the scaling to heat exchangers with more tubes is not linear at higher pressure differentials, likely due to the geometry which leads to enhanced forced convection in the innermost region of the multi-tube pattern for large heat exchangers. In any case, the DS"~20k design approach will be conservative and the envelope of both curves will be used as a lower limit of the required number of tubes. Depending on the available pressure difference at the two-phase heat exchanger in DS"~20k, accounting for all pressure drops in the system, this extrapolation will serve as design basis for the required exchange surface.

\subsection{Condenser icing test}
\label{subsec:SystemExtremes}
\begin{figure}[t]
\centering
\includegraphics*[width=1\textwidth]{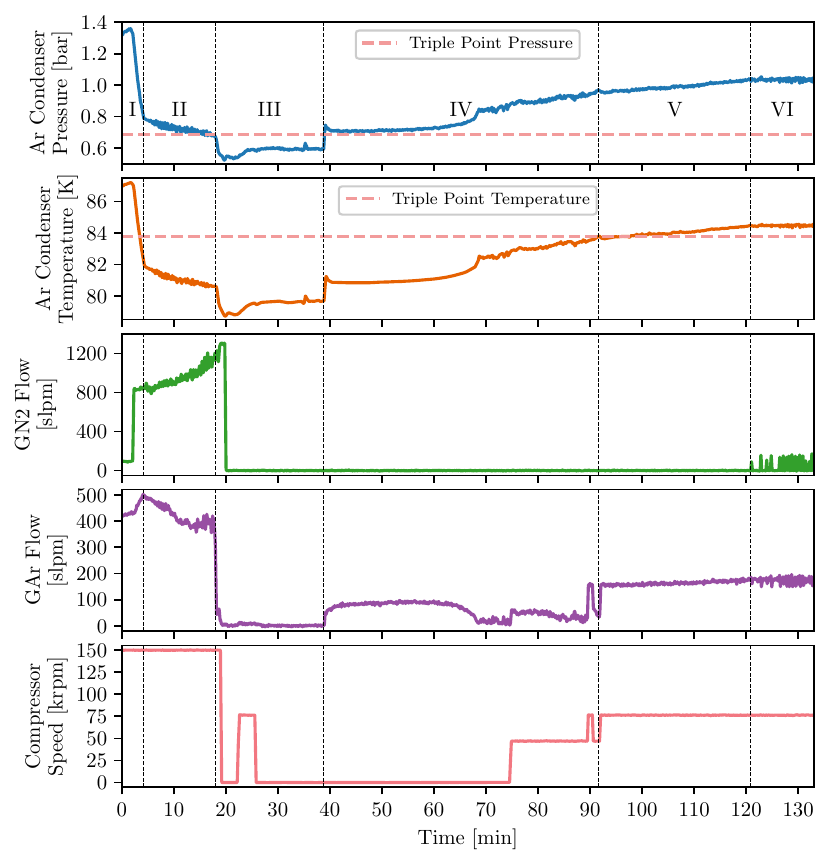}
\caption[]{Condenser icing test data. See text for details. The temperature sensor is attached to the condenser body downstream, not to the heat transfer surface.}
\label{fig:CondenserIcing}
\end{figure}

As mentioned in the preceding sections, the condenser is operated with LN$_2$ at a saturation temperature lower than the argon melting temperature. However, the condenser is not flooded with LN$_2$. Instead, LN$_2$ is dosed in such a way that the argon does not freeze on the inner surface of the condenser tubes in normal operations but rather stays $\SI{0.6}{K}$ above the triple point, see figure~\ref{fig:LongtermStability}. In order to assess the risk of argon solidification on the heat exchange surface of the condenser during phases of high cooling power demand, we have performed a dedicated test and intentionally formed solid argon~(SAr) on the condenser. To this end, starting from normal operations in recirculation mode, we have raised the cryostat pressure to $\SI{\sim 1.3}{bara}$, similar to the procedure described in section~\ref{subsec:CoolingpPerformancePressureStability}, and have then abruptly reduced the cryostat pressure setpoint to $\SI{0.7}{bara}$, leading momentarily to maximum cooling power. The slow monitoring data of the test is shown in figure~\ref{fig:CondenserIcing}. The phases and events marked in the plot are explained below.

\begin{enumerate}[label=\Roman*,itemsep=0pt]
    \item{Cooling start: starting at $\SI{\sim 1.3}{bara}$, the cooling control flow valve FV~PIDN is fully opened. Within the first $\SI{2.5}{min}$, $\SI{11.5}{L}$ LN$_2$ are flushed into the condenser as can be derived from the phase separator liquid level. Only $\SI{3.2}{L}$ are evaporated in this period as can be integrated from the GN$_2$ flow. Thus, the nitrogen side of the condenser, which has a volume of $\SI{\sim 8.4}{L}$, was completely filled with LN$_2$ at this point. The temperature falls rapidly below the triple point, indicating a transition from the liquid to the solid regime on the condenser tubes.}
    \item{Ice formation and outlet clogging: a sudden change in the argon temperature and pressure slope indicates a reduced heat transfer, likely due to the reduction of exchange surface and a reduced heat transfer coefficient due to the formation of a thick SAr layer on the tubes. In this phase, the condenser outlet of the argon side gets clogged. The GN$_2$ flow still rises, even though the pressure differential across the cooling line is observed to decrease (not shown in figure~\ref{fig:CondenserIcing}). This is due to the falling temperature, and thus increasing mass density, of the GN$_2$ in the heat exchangers downstream. The ripples in the GN$_2$ flow and condenser condition data are due to the start of the refilling of the phase separator which is trying to maintain a constant liquid level. The transfer line upstream of the phase separator is still warm, however, delivering only GN$_2$ at this point. LAr is now present on both argon channels of the heat exchanger cascade.}
    \item{Inlet clogging and cooling stop: the condenser inlet of the argon side is clogged, visible as a sudden drop in GAr flow at constant compressor speed. The pull from the condenser, acting as a cryogenic pump, is suddenly removed. The condenser pressure falls below the triple point. The compressor is turned off by operator action, a later attempt to restart during this phase results in very little flow. Shortly after, the LN$_2$ supply is stopped by raising the cryostat setpoint to its nominal value. This is visible as a sharp decrease in GN$_2$ flow due to the closing of the cooling control valve FV~PIDN. The gas outlet of the phase separator is now acting as a condenser vent, visible as an increase in the opening of the pressure control valve~PV~PS. A total of $\SI{\sim 27}{L}$ LN$_2$ has been consumed up until this point, capable of removing a total of $\SI{8.5}{MJ}$ over the course of the test\footnote{The GN$_2$ outlet downstream of the heat exchanger cascade is progressively cooling with a minimum at $\SI{255}{K}$. For the calculation, we have used the GN$_2$ enthalpy at $\SI{278}{K}$, which is the average temperature of the GN$_2$ that is leaving the condenser box during the test.}. This energy is sufficient to solidify $\SI{33.4}{kg}$ of argon when starting with vapour at $\SI{200}{K}$ and $\SI{1}{bara}$. The volume of this amount of SAr exceeds the $\SI{\sim 5}{L}$ volume of the argon side of the condenser by a factor of 3. In lack of available heat transfer surface, it is thus unlikely that this amount of SAr is built but plausible that the SAr extends into the connected piping. Now effectively separated from the condenser volume, it is observed that the thermodynamic conditions upstream of the condenser, with the temperature slightly above and the pressure significantly below the triple point indicate the presence of vapour only in the heat exchangers for the first two-thirds of this phase (not shown in figure~\ref{fig:CondenserIcing}).}
    \item{Inlet de-icing and warming: the condenser inlet de-ices, and the pressures of the heat exchanger cascade and the condenser equalise. The condenser is still filled with SAr. The pressure crosses the triple point. The local underpressure created by the ice block brings new gas to the condenser, visible as non-vanishing GAr flows with the compressor off. The SAr is warming but not melting yet.}
    \item{Condenser (-outlet) de-icing and recovery: $\SI{72}{min}$ after the cooling control valve FV~PIDN was closed, the argon temperature at the condenser outlet crosses the triple point, aided by increasing the compressor speed. The condenser and its outlet is de-iced. }
    \item{Stable operations: the cryostat pressure reaches its operating point and the cooling restarts, bringing the system to stable operating conditions. From the moment the cooling control valve FV~PIDN was closed, the system needed $\SI{101}{min}$ to recover completely.}
\end{enumerate}

It is noted that the performed test has brought the condenser into a thermodynamic regime which is far from the conditions present during normal operations -- even during phases of high demand for cooling power. In the case of high-speed gas filling at $\SI{\sim 1000}{slpm}$ in DS"~20k, the latent heat exchange of the condensing argon requires a similar cooling power but the injected GAr adds enough heat to the system to maintain a stable pressure and temperature. In this test, a sudden enormous deviation of $\SI{0.6}{bard}$ of the pressure setpoint from the actual cryostat pressure was an anomalous event for the cooling control, which generally works in a continuous way, i.e.~the cooling power responds slowly to a physical steady pressure increase in the cryostat. This has led to an enormous instant cooling power which could not be absorbed by argon condensation only at the system's heat load. The system has, however, behaved predictably and has recovered to its normal state without the operator's intervention after restoring the nominal cryostat pressure setpoint (compressor speed changes were not necessary actions). None of the parameters has exceeded the working range and no safety relief valve has opened. As intended, we have been able to create SAr on the condenser. However, not only the condenser outlet has been clogged by SAr but also the inlet, hence separating the cryostat from the condenser. Even though only possible during anomalous operations when the cooling power exceeds the latent heat that can be removed from the argon flow, this presents a potential risk of losing UAr in DS"~20k in case of rising cryostat pressure. We discuss mitigation strategies in section~\ref{sec:DS-20kOutlook}. We note however already here that due to the much higher thermal momentum and gas ullage of the UAr reservoir in the DS"~20k vessel, neglecting the cooling provided by the AAr bath, the time scale at which the pressure in the vessel changes is expected to be significantly larger than the one of our test cryostat, leaving sufficient time for the condenser to de-ice.   

\section{Outlook to the DarkSide-20k system and conclusion}
\label{sec:DS-20kOutlook}

\subsection{Implications of the tests}
\label{subsec:Implications}

The commissioning tests and benchmarking measurements of the previous section yield valuable input for the finalisation of the DS"~20k system design. During the tests, the system successfully demonstrated its power, efficiency and stability, but also revealed room for improvement in certain aspects. These points are addressed by modifications to fulfil the requirements set in section~\ref{sec:RequirementsConceptDesign}. Those are listed below:
\begin{itemize}[itemsep=0pt]
    \item{Compressor: the Celeroton CT"~1000"~Ar compressor~\cite{Celeroton} performs to specifications and has been satisfactory for the needs of the test bed. However, it is likely that a maximum compression ratio of 1.6 at $\SI{500}{slpm}$ is insufficient for the DS"~20k system. While the flow can be increased using two pumps in parallel, the compression of a single stage does not overcome the total predicted pressure drop of the loop which is $\SI{\sim 1}{bard}$\footnote{The total pressure drop of the loop at $\SI{1000}{slpm}$ is dominated by the Entegris MegaTorr\textsuperscript{\textregistered} PS5"~MGT200"~R gas purifier~\cite{Getter} ($<\SI{300}{mbard}$), the heat exchanger cascade (target: $\SI{300}{mbard}$), the radon trap ($<\SI{100}{mbard}$, see section~\ref{subsec:OngoingActivitites}) and the two-phase heat exchanger ($\SI{300}{mbard}$).}. This compressor was not tested for in-series operation of several compressor stages. The final choice of the recirculation pump is pending at the time of writing and other models are being considered, see section~\ref{subsec:OngoingActivitites}.}
    \item{Heat exchanger cascade: even though measured at room temperature, and thus at lower argon mass density, the flow resistance results imply that the pressure drop over the heat exchanger cascade is too high. The modules will therefore be replaced by appropriate substitutes. A model to describe the pressure drop has been presented in section~\ref{subsec:CompressorPerformanceCircuitResistance} and will be used to update the design while maintaining the high system efficiency. The first heat exchanger in the chain~(HE1), in the direction of the return flow, will be dimensioned to allow for two-phase heat exchange at the target fill speed. This will be relevant for direct LAr filling of the system from transport and storage containers and simultaneous gas purification of the injected argon.}
    \item{Cooling control: the cooling control unit has performed as expected in terms of stability and flow. The flow coefficient of the PID-controlled valve FV~PIDN allows for high flow rates but is impractical for small flows, resulting in abrupt changes of the flow at minimal changes of the orifice. To avoid this, a control valve with a smaller flow coefficient will be installed in parallel which is capable of tuning the flow in the regime of $\SIrange{0}{100}{slpm}$ in GN$_2$ flows. This is necessary to control the system if the detector instrumentation is off. The GN$_2$ flow expected to compensate the heat load from instrumentation is $\SIrange{200}{300}{slpm}$, which is well within the controllable range of the presently installed valve. A fourth bypass will be added consisting of a manual needle and isolation valve for manual tuning. While in the test bed, the GAr and the GN$_{2}$ volumes inside the bellow valve are separated by a single edge-welded bellow, the design for DS"~20k will be upgraded with an additional stainless steel bellow. Although the bellow is virtually not actuated in phases of constant UAr vessel pressure, this double containment provides an additional layer of safety against contamination or loss of the UAr in case of fatigue failure. It also allows for monitoring of the pressure in the intermediate space, reducing the risk of undetected leakage through a bellow. Additionally, the bellow valve design will be upgraded with metal gaskets.}
    \item{Two-phase heat exchanger: the test data is the design basis for dimensioning the final tube heat exchanger. To keep the size of the heat exchanger acceptable with a diameter of $\SI{\sim 750}{mm}$, the target operating point is set to $\SI{300}{mbard}$, yielding less than $1200$~tubes. Depending on the performance of the chosen recirculation pump and the total pressure drop of the loop, this target might get altered in the future.}
    \item{Condenser: to give the operator the ability to actively de-ice the condenser requires the outer body to be equipped with an electric belt heater. This element might also be integrated into the functional logic of the system to enable de-icing automatically based on the temperature and pressure conditions in the condenser. Further precautions include an interlock that reduces the cooling power in case of condenser icing and an alarm sent to the operator in such an event. The detector control is further discussed in section~\ref{subsec:OngoingActivitites}. Besides its function as a defrost element, the heater can also compensate for a potential remnant cooling power at zero GN$_2$ flow due to small spills of LN$_2$ from the chicken feeder into the condenser arising from pressure fluctuations in the phase separator. As noted in section~\ref{subsec:CoolingpPerformancePressureStability}, this effect is expected to be eliminated in the DS"~20k system due to the alternating use of two nitrogen phase separators.} 
\end{itemize}

\subsection{Ongoing design and testing activities}
\label{subsec:OngoingActivitites}
Efforts towards the finalisation of the design of the DS"~20k UAr cryogenics system are ongoing. These include the analysis of the operating modes and corresponding modifications of the engineering P\&ID, studies of the argon distribution in the stainless steel vessel, fluid flow and thermal simulations of the inner detector, in particular of the cool-down of detector components, as well as subsystem design efforts. We highlight a few more activities below. 

\begin{figure}[t]
\centering
\begin{subfigure}[b]{0.44\textwidth}
\centering
\includegraphics*[height=0.2\textheight]{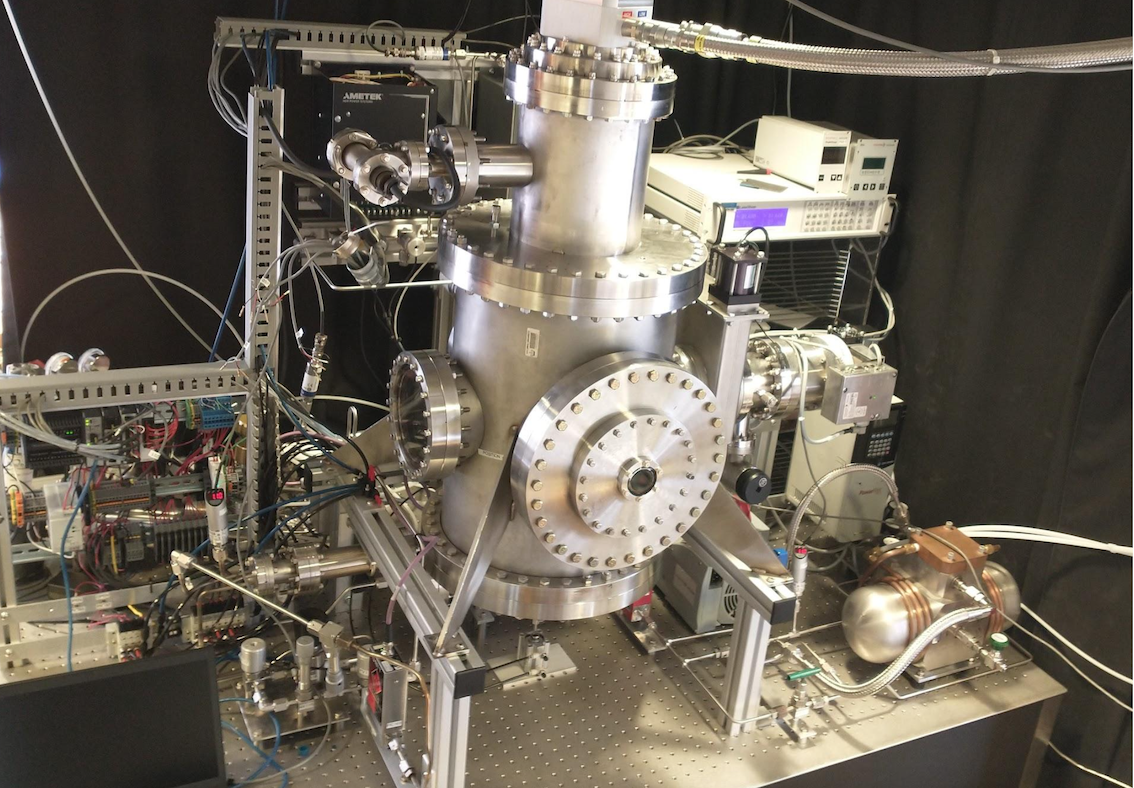}
\end{subfigure}
\quad
\begin{subfigure}[b]{0.52\textwidth}
\centering
\includegraphics*[height=0.2\textheight]{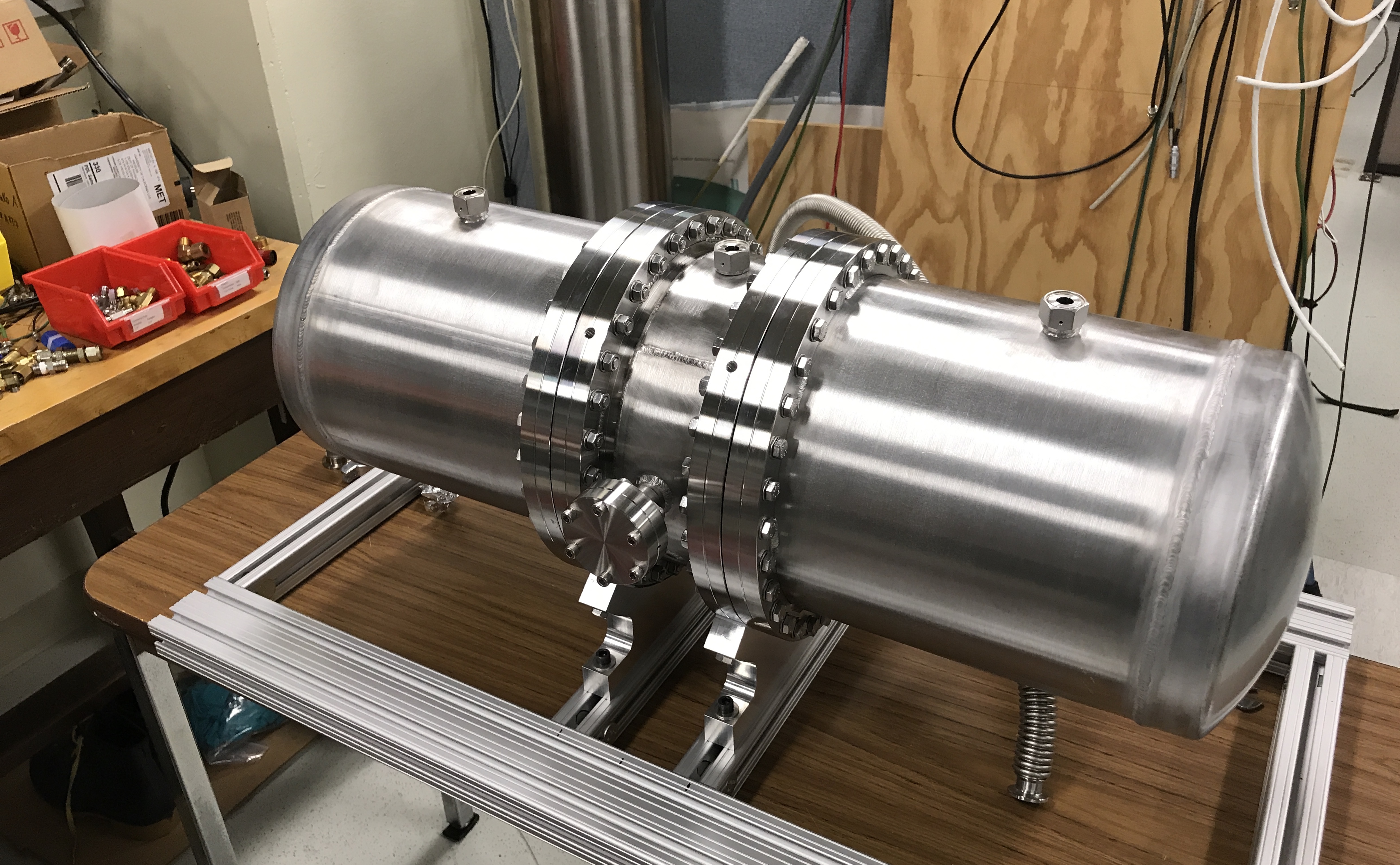}
\end{subfigure}
\caption[]{(Left): Test bench for radon adsorption measurement at Columbia University. (Right): In-house developed candidate recirculation pump at the University of California, Los Angeles.}
\label{fig:RnSetupLiHanPump}
\end{figure}

\paragraph{Radon trap}
As mentioned in section~\ref{subsec:CondenserBox}, for DS"~20k the condenser box will be upgraded with a charcoal-based radon trap between HE~1 and HE~2 and a downstream particulate filter. Additional lines upstream and downstream of the radon trap and an additional isolation valve will allow the regeneration of the charcoal with heat while the recirculation loop is active. The test of a high-purity activated charcoal sample (spheres with $\SI{0.4}{mm}$ diameter) from TALAMON~\cite{Talamon} (formerly Saratech from BLÜCHER~\cite{Bluecher}) is ongoing. Compared to alternatives such as from Carbo-Act~\cite{CarboAct}, the TALAMON product is considerably lower in cost and features a much narrower size distribution of the charcoal beads. This reduces the flow resistance uncertainty, at minimally lower radon removal capability and slightly increased radioactivity~\cite{Pushkin:2018wdl,Chen:2021hva}. To evaluate the properties of the sample in hand, measurements regarding the temperature- and flow-dependent radon dynamic adsorption coefficient (see figure~\ref{fig:RnSetupLiHanPump}~(left)), the specific radiogenic activity and the flow resistance of various cylindrical trap geometries are conducted. The goal of these measurements is the design of a trap with a statistically predictable radon removal and known flow resistance which is minimised as much as reasonably achievable (below $\SI{100}{mbard}$). The sizing of the trap can then be decided based on further inputs regarding the maximum allowed radon activity in the TPC, the expected activity at the radon trap inlet, the target flow rate and the expected maximum radon trap temperature.  

\paragraph{Recirculation pump}
Based on the successful development of the DS"~50 recirculation pump, an improved DS"~20k candidate pump was designed and fabricated in-house (see figure~\ref{fig:RnSetupLiHanPump}~(right)). The positive displacement pump has a $\SI{500}{slpm}$ design flow and a target compression factor of $2.5$. Two pumps can be connected in parallel to reach the DS"~20k design flow rate. The pump features two $\SI{1}{kW}$ linear motors from Lihan Cryogenics~\cite{Lihan} with a maximum individual stroke of $\SI{10}{mm}$ and a piston diameter of $\SI{120}{mm}$. The motors are placed face-to-face, aligned by a central body through which the fluid is passing, and operated at $\SI{50}{Hz}$ in opposite phases to counteract vibrations. Custom laser-cut reed valves made from Alleima\textsuperscript{\textregistered} (Sandvik) 7C27Mo2 compressor valve steel~\cite{Alleima}, one for each motor, determine the gas flow direction. A vibration sensor can detect over-stroke or abnormal vibration. An integrated water cooling system removes heat from the motors and the central compression chamber. The pump performance is currently being evaluated. 

\paragraph{Slow monitoring and control}

\begin{figure}[t]
\centering
\includegraphics[width=0.618\textwidth]{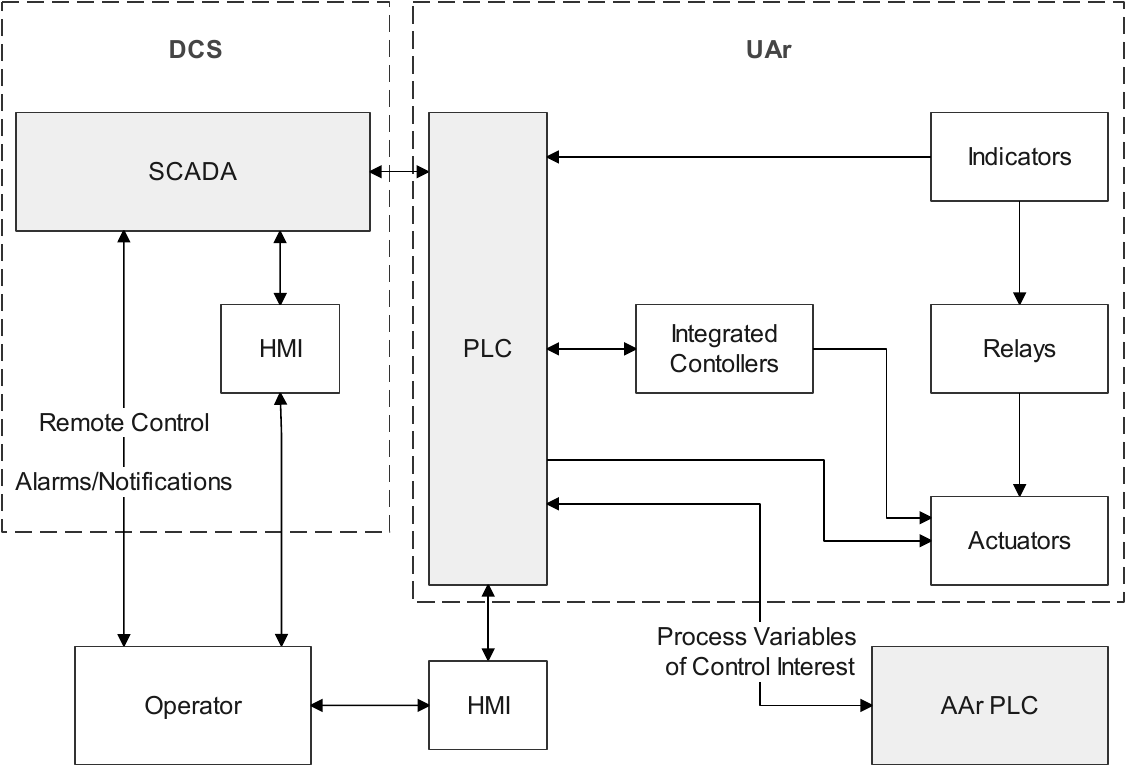}
\caption[]{Schematic of the UAr slow monitoring/control building blocks.}
\label{fig:DS-20kSlowMonitoring}
\end{figure}

The slow monitoring, control and safety system is currently under development. We provide here a high-level conceptual overview of the aspects that concern the UAr cryogenics system. A schematic of the planned system is shown in figure~\ref{fig:DS-20kSlowMonitoring}. The UAr cryogenics system's components will be read or controlled via a dedicated programmable logic controller~(PLC), which will run the UAr cryogenics functional logic that does not require decision blocks from other systems. Equipment such as sensors will be hard-wired connections, in particular, those for the exchange of control-critical process variables to or from other systems such as the AAr cryogenics. The higher-level DS"~20k detector control system~(DCS) will use a WinCC Open Architecture-based supervisory control and data acquisition system~(SCADA)~\cite{WINCCOA}. The interface with the UAr cryogenics PLC is implemented at the SCADA level. It is also at this level that the operator can interface with all other systems, in particular via human machine interfaces~(HMI), and e.g.~change setpoints or actuate valves. The SCADA is the instance from which alarms and notifications are sent to the operator.

The UAr cryogenics functional logic comprises two types of controls: control loops and interlocks. The former can be implemented either on the PLC level or in integrated equipment. Control loops such as pressure-based cooling, are usually continuously running but may only be active for certain operations. Interlocks such as in the event of condenser icing, discussed in section~\ref{subsec:Implications}, are usually always active, especially if they are relevant for safety. Some interlocks may be specific to the operation and may be armed or disarmed by the operator.

In parallel to the DCS, there will be an independent detector safety system~(DSS) which interacts directly with the detector equipment, like from the UAr cryogenics. It includes hard-wired connections of safety-critical components and related process variables such as present between the UAr and the AAr cryogenics (not shown in figure~\ref{fig:DS-20kSlowMonitoring}). Unlike the DCS, which handles normal operations, the function of the DSS is to detect abnormal and potentially harmful conditions at a higher level of safety, with the goal of minimising the resulting risk to personnel and damage to the equipment by taking protective actions. 

\subsection{Conclusion}
\label{subsec:Conclusion}
This paper provides an overview of the design, testing and development status of the UAr cryogenics system for DS"~20k, a next-generation WIMP detector deploying $\SI{\sim 100}{t}$ of UAr and whose construction at LNGS is underway. The core components of the system have been successfully commissioned and benchmarked in a test bed at CERN and LNGS.

The system contains a gas purification loop which must be optimised for high argon flow rates, allowing $\SI{1000}{slpm}$. For this reason, we have tested a radial compressor with gas bearing as a candidate recirculation pump for DS"~20k and found it to perform according to specifications. The compression ratio of the tested model is slightly above $1.6$ at $\SI{500}{slpm}$. Even though two compressors could be used in parallel to reach the required flow, the compression is insufficient for the expected circuit resistance which requires a minimum of $2.0$ at $\SI{1000}{slpm}$. The performance of another promising candidate recirculation pump, developed and manufactured in-house, is currently being tested. It was further found that the flow resistance of the gas heat exchanger cascade, which is necessary for high energy use efficiency during recirculation, is not acceptable. These heat exchangers will be replaced by other modules for DS"~20k. The presently installed heat exchangers however, account for an efficient design with a dynamic heat load recovery of $\SI{95}{\%}$, tested up to more than $\SI{220}{slpm}$. This allows the operation at $\SI{1000}{slpm}$ argon flow with only $46\pm2\,\si{slpm}$ nitrogen consumption compared to the zero argon flow demand, assuming that the linear relation of the GN$_2$ and the GAr flow is preserved for higher GAr flow rates. High energy use efficiency is an important requirement for DS"~20k due to its long runtime of more than $\SI{10}{years}$. Cooling is provided to the system with an LN$_2$-operated condenser and controlled based on the pressure of the argon reservoir. Deploying LN$_2$ with a saturation temperature below the triple point of argon, the condenser uses a self-regulating dosing method, to ensure stable operations with LAr. We have thoroughly tested this design and operated it intentionally also outside standard operating conditions, demonstrating its performance and confirming its working principle. Modelling and tests show that the cooling system can fulfil the $\SI{8}{kW}$ cooling power requirement. The redundant and power failure-immune cooling control provides stable operating conditions while being highly responsive to changes in the argon reservoir pressure. In between nitrogen phase separator refills, the test cryostat pressure was stable within $\SI{0.2}{mbar}$ RMS using the PID-controlled valve, for normal operations, and within $\SI{0.1}{mbar}$ RMS using the bellow valve, in power-failure mode, for the GN$_{2}$ flow control.
Based on these observations, it is expected that the pressure stability requirement of $\SI{1}{mbar}$ RMS can be fulfilled in DS"~20k due to the following differences: the alternating use of two N$_2$~phase separators, the larger available gas ullage and the use of a control valve for the cooling control with a flow-coefficient which is appropriately sized for the respective GN$_2$ flow regime. This is part of the set of improvements and upgrades that will be made to the system design before its integration into the DS"~20k facility. Before this work, the test bed will service the DS"~20k Mockup TPC campaign.  

Ongoing testing activities include the adsorption capability and radiogenic activity of an activated charcoal sample for the radon trap and flow resistance measurements of various trap geometries. The final design of the DS"~20k purification loop, the procurement of its components, and the development of the functional logic as well as the monitoring and control infrastructure, are in progress.

\acknowledgments{The collaboration acknowledges the support and supplies of the CERN Cryolab and would like to thank in particular J.~Bremer and T.~Koettig for hosting the RE37 experiment. We further thank R.~Vasseur for welding the internals of the condenser box. We also thank LNGS for hosting our setup underground and are grateful to all administrative and technical staff that fostered the testing campaign, in particular to F.~Di Eusanio, F.~Tacca and M.~Balata. We thank F.~Pietropaolo for providing the laboratory infrastructure and the liquid argon for the small-scale tube heat exchanger measurements. We acknowledge active discussions with R.~Pengo in the context of the INFN review panel. 

This report is based upon work supported by the U.~S.~National Science Foundation (NSF) (Grants No.~PHY-0919363, No.~PHY-1004054, No.~PHY-1004072, No.~PHY-1242585, No.~PHY-1314483, No.~PHY-1314507, No.~PHY-1622337, No.~PHY-1812482, No.~PHY-1812547, No.~PHY-2310091, No.~PHY-2310046, associated collaborative grants, No.~PHY-1211308, No.~PHY-1314501, No.~PHY-1455351 and No.~PHY-1606912, as well as Major Research Instrumentation Grant No.~MRI-1429544), the Italian Istituto Nazionale di Fisica Nucleare (Grants from Italian Ministero dell’Istruzione, Università, e Ricerca Progetto Premiale 2013 and Commissione Scientific Nazionale II), the Natural Sciences and Engineering Research Council of Canada, SNOLAB, and the Arthur B.~McDonald Canadian Astroparticle Physics Research Institute. 
This work received support from the French government under the France 2030 investment plan, as part of the Excellence Initiative of Aix-Marseille University -- A*MIDEX (AMX-19-IET-008 -- IPhU).
We also acknowledge the financial support by LabEx UnivEarthS (ANR-10-LABX-0023 and ANR18-IDEX-0001), Chinese Academy of Sciences (113111KYSB20210030) and National Natural Science Foundation of China (12020101004).
This work has been supported by the S\~{a}o Paulo Research Foundation (FAPESP) grant 2021/11489-7
and by the National Council for Scientific and Technological Development (CNPq).
Support is acknowledged by the Deutsche Forschungsgemeinschaft (DFG, German Research Foundation) under Germany's Excellence Strategy -- EXC 2121: Quantum Universe -- 390833306.
The authors were also supported by the Spanish Ministry of Science and Innovation (MICINN) through the grant PID2022-138357NB-C22, the ``Atraccion de Talento'' grant 2018-T2/TIC-10494, the Polish NCN (Grant No.~UMO-2019/33/B/ST2/02884), the Polish Ministry of Science and Higher Education (MNiSW, grant number 6811/IA/SP/2018), the International Research Agenda Programme AstroCeNT (Grant No.~MAB/2018/7) funded by the Foundation for Polish Science from the European Regional Development Fund, the European Union’s Horizon 2020 research and innovation program under grant agreement No 952480 (DarkWave), the Science and Technology Facilities Council, part of the United Kingdom Research and Innovation, and The Royal Society (United Kingdom), and IN2P3-COPIN consortium (Grant No.~20-152). We also wish to acknowledge the support from Pacific Northwest National Laboratory, which is operated by Battelle for the U.~S.~Department of Energy under Contract No.~DE-AC05-76RL01830.
This research was supported by the Fermi National Accelerator Laboratory (Fermilab), a U.~S.~Department of Energy, Office of Science, HEP User Facility. Fermilab is managed by Fermi Research Alliance, LLC (FRA), acting under Contract No.~DE-AC02-07CH11359.
}

\newpage
\appendix
\section{P\&ID Symbols}
\label{app:PIDSymbols}

\begin{table}[h]
\footnotesize
\caption{Legend of the symbols used in the P\&ID in figure~\ref{fig:UArCryoTestPID}.}
\label{tab:PID_legend}
\begin{minipage}[t]{0.49\textwidth}
\setlength{\tabcolsep}{3pt}
\renewcommand{\arraystretch}{0.98}
\begin{tabularx}{\textwidth}{| >{\setlength{\baselineskip}{0.9\baselineskip}}c | >{\setlength{\baselineskip}{0.9\baselineskip}}c | >{\setlength{\baselineskip}{0.9\baselineskip}}X |}
\hline
\bf{Specif.}  & \bf{Identif.}  & \bf{Component} \\
\hline
AV      &  01--14  & Angle Valve 01--14\\
\hline
B      &    & Bottle\\
    &  Ar  & Argon\\
    &  He  & Helium\\
\hline
BV &    & Buffer Volume\\
    &  PD  & Pump Downstream\\
    &  PU  & Pump Upstream\\
\hline
C &    & Cryostat\\
\hline
CD &  1/2  & Condenser 1/2\\
\hline
FIT &  & Flow Indicator Transmitter\\
    & Ar & Argon\\
    & N & Nitrogen\\
    \hline
FM & C & Filter Mesh Cryostat\\    
\hline
FV &  & Flow-control Valve\\
    & PN & Pneumatic Nitrogen\\
    & PIDN & Proportional-Integral-Derivative Nitrogen\\
    \hline
GCP &   & Gas Compressor Pump\\
\hline
HE &    & Heat Exchanger\\
    &  1--5  & 1--5\\
    &  W  & Water\\
\hline
HS & N   & Heat Sink Nitrogen\\
\hline
HV &    & Hand Valve\\
    &  BAr  & Bottle Argon\\
    &  BHe  & Bottle Helium\\
    &  BV  & Buffer Volume\\
    &  C  & Cryostat\\
    &  He  & Helium\\
    &  HeV  & Helium Vent\\
    &  LT1/2  & Level Transmitter\\
    &  PC  & Pressure Cryostat\\
    \hline
LT & PS & Level Transmitter Phase Separator\\
\hline
LV & PS & Level-control Valve Phase Separator\\
\hline
PI &  C  & Pressure Indicator Cryostat\\
\hline
PIH &   & Pressure Indicator High\\
    &  Ar & Argon\\
    &  He & Helium\\
    \hline
PIL &   & Pressure Indicator Low\\
    &  Ar & Argon\\
    &  He & Helium\\
    \hline
\end{tabularx}
\end{minipage}
\hfill
\begin{minipage}[t]{0.49\textwidth}
\setlength{\tabcolsep}{3pt}
\renewcommand{\arraystretch}{0.98}
\begin{tabularx}{\textwidth}{| >{\setlength{\baselineskip}{0.9\baselineskip}}c | >{\setlength{\baselineskip}{0.9\baselineskip}}c | >{\setlength{\baselineskip}{0.9\baselineskip}}X |}
\hline
\bf{Specif.}  & \bf{Identif.}  & \bf{Component} \\
\hline
PIT &    & Pressure Indicator Transmitter\\
    &  A1/2  & Argon Loop 1/2\\
    &  C1/2  & Cryostat 1/2\\
    &  CDA  & Condenser Argon side\\
    &  CDN  & Condenser Nitrogen side\\
    &  He  & Helium\\
    &  N1/2  & Nitrogen 1/2\\
    &  PD  & Pump Downstream\\
    &  PU  & Pump Upstream\\
    \hline
PR &    & Pressure Regulator\\
&  Ar  & Argon\\
&  He  & Helium\\
\hline
PS &  & Phase Separator\\
\hline
PT &  PS  & Pressure Transmitter Phase Separator\\
\hline
PV &  & Pressure-control Valve \\
    & He & Helium\\
    & PS & Phase Separator\\
    \hline
RD &  PS  & Rupture Disk Phase Separator\\
\hline
RV &    & Relief Valve\\
    &  C1/2  & Cryostat 1/2\\
    &  CDA  & Condenser Argon side\\
    &  CDN  & Condenser Nitrogen side\\
    &  He  & Helium\\
    &  N  & Nitrogen\\
    & NS & Nitrogen Supply\\
    &  PA  & Pump Argon\\
    &  PS  & Phase Separator\\
    &  V1/2  & Vacuum 1/2\\
    \hline
TT &   & Temperature Transmitter\\
    \hline
VIT &    & Vacuum Indicator Transmitter\\
    &  C  & Cryostat\\
    & CB   & Condenser Box\\
    \hline
VPR    &  1--3  & Vacuum Pump Roughing 1--3\\
\hline
VPT    &  1/2  & Vacuum Pump Turbomolecular 1/2\\
\hline
XV &  & ON/OFF Valve\\
    &  A1--8  & Argon 1--8\\
    & N & Nitrogen\\
    \hline
\end{tabularx}
\end{minipage}
\end{table}
\section{Multi-tube heat exchanger for phase change}
\label{app:MultiTubeGeometry}
The multi-tube heat exchanger geometry is deployed for phase change in two locations within the system: as a nitrogen-argon condenser in the condenser box and as a two-phase argon-argon heat exchanger in the test cryostat. Below, we present the modelling of the maximum possible cooling power of the first configuration, followed by an analysis of the heat transfer of the second. The prediction of the argon two-phase heat transfer capability of the multi-tube geometry is compared to a small-scale measurement with 19~tubes that has been performed before deploying the 127"~tube heat exchanger in the test cryostat.

\subsection{Modelling of nitrogen-argon and argon-argon heat transfer}
\label{appsubsec:ModellingTubeGeometry}

This section aims to model the expected heat transfer in the LN$_{2}$"~GAr tube condenser and the LAr"~GAr tube heat exchanger. Generally, the heat transfer~$\dot Q$ through a layer $i$ of material, or gas/liquid-solid boundary is expressed as a function of the temperature difference~$\Delta T_i$ across that layer, the corresponding heat transfer coefficient~$h_i$, and the layer's cross-sectional area  $A_i$ using Newton's law of cooling:
\begin{align} \label{film}
    \dot Q = h_i\cdot A_i \cdot \Delta T_i \quad.
\end{align}
For initial modelling purposes, we consider a single-unit stainless steel tube geometry, which can be later scaled to the total number of tubes used. This tube is characterised by its outer diameter, wall thickness~$d_\mathrm{Wall}$, and length~$L$. We will focus on the shell surface area, neglecting the contributions from the top and bottom circular areas of the tube.

For the LN$_{2}$"~GAr condenser, we aim to determine the maximum achievable heat transfer before argon freezes on the inner tube surface. For that reason, the tube's inner surface temperature is fixed at the argon triple point ($\SI{83.81}{K}$~\cite{NIST}), representing the minimum temperature before solidification. We assume that a constant and uniform temperature across the tube's outer surface can be maintained to achieve this temperature boundary condition on the tube's inner surface. To estimate the heat transfer coefficient between the argon and the stainless steel tube's inner surface, we assume film condensation occurs on a flat vertical surface, rather than in a circular tube. This is valid because the tube diameter significantly exceeds the expected liquid argon film thickness. Following reference~\cite{Bergman} (section 10.7), the average heat transfer coefficient~$h_\mathrm{GAr-SS}$ for a vertical surface of length~$L$ with a temperature difference between the saturated argon vapor~$T_\mathrm{GAr,sat}$ and the tube's inner stainless steel surface $T_\mathrm{S,in}$ is given by
\begin{align}  \label{eq:film2}                              
h_\mathrm{GAr-SS} &= 0.943 \left(\frac{g\cdot \rho_\mathrm{LAr} (\rho_\mathrm{LAr}-\rho_\mathrm{GAr}) k_\mathrm{LAr}^3\cdot H^\mathrm{Ar}_{\text{vap}}}{\mu_\mathrm{LAr}\cdot (T_\mathrm{GAr,sat} - T_\mathrm{S,in})\cdot L}\right)^{1/4} \quad,
\end{align}   
with $\rho_\mathrm{LAr}$ being the liquid density, $\rho_\mathrm{GAr}$ the gaseous density, $k_\mathrm{LAr}$ the liquid thermal conductivity, $\mu_\mathrm{LAr}$ the dynamic liquid viscosity, $H^\mathrm{Ar}_{\text{vap}}$ the latent heat of vaporisation of argon, and $g$ the gravitational acceleration.
The LAr properties are evaluated at the film's temperature $T_\mathrm{film,in} = (T_\mathrm{GAr,sat} + T_\mathrm{S,in}) / 2$~\cite{NIST}.

\begin{figure}[t]
\centering
\includegraphics[width=0.618\textwidth]{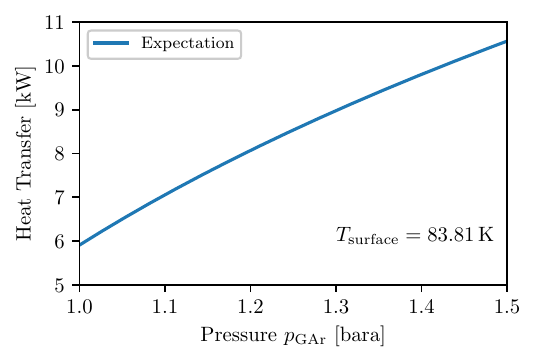}
\caption[]{Maximum possible heat transfer on the condenser tubes for a 127-tube condenser, assuming film condensation, as a function of the argon pressure in the tubes. The temperature boundary of the argon-wetted tube surface is fixed to the argon triple point to maximise the heat transfer. In DS"~20k the condenser will be operated with an argon pressure of $\SIrange{1}{1.1}{bara}$.}
\label{fig:CondenserModelMaxCoolingPower}
\end{figure}

The saturated argon temperature $T_\mathrm{GAr,sat}$, and thus the heat transfer coefficient $h_\mathrm{GAr-SS}$, is pressure dependent~\cite{NIST}. To investigate the influence of the operational argon pressure on the heat transfer performance, we computed equation~\eqref{eq:film2} for various operating pressures $p_\mathrm{GAr}$ on the argon side. Subsequently, we scaled the results to account for the actual number of tubes within the condenser. This procedure allowed us to estimate the achievable heat transfer capacity of the condenser as a function of argon pressure. The results are presented in figure~\ref{fig:CondenserModelMaxCoolingPower}. For instance, at a pressure of $\SI{1.2}{bara}$, the estimated heat transfer for the 127"~tube condenser is approximately $\SI{8}{kW}$. Note that this result represents an upper limit to the possible cooling power of the condenser.

Similar to the LN$_{2}$"~GAr condenser, the LAr"~GAr tube heat exchanger utilises argon on the inside of the tube at an operation pressure~$p_\mathrm{GAr}$, and corresponding saturated temperature~$T_\mathrm{GAr,sat}$. However, the tube's exterior is surrounded by LAr at a distinct pressure~$p_\mathrm{LAr}$ and corresponding temperature~$T_\mathrm{LAr,sat}$. This configuration necessitates considering three heat transfer mechanisms: heat transfer from the inner GAr to the stainless steel wall, heat conduction across the tube's wall, and heat transfer from the outer stainless steel wall to the LAr.

To calculate the heat transfer coefficient~$h_\mathrm{ss}$ through the stainless steel wall with inner surface temperature~$T_\mathrm{S,in}$ and outer surface temperature $T_\mathrm{S,out}$, we can utilise its thermal conductivity~$\lambda_\mathrm{SS}$ and thickness~$d_\mathrm{Wall}$:
\begin{align}
     h_\mathrm{SS} &= \frac{\lambda_\mathrm{SS}}{d_\mathrm{Wall}} \label{eq:hss} \quad.
\end{align}

Fluid properties and flow regimes are often characterised by dimensionless numbers. In this analysis, we employ the Reynolds number~$Re$, the Grashof number~$Gr$, and the Prandtl number~$Pr$ to describe the flow of LAr. Additionally, the Rayleigh number~$Ra$, defined as the product of~$Gr$ and~$Pr$~\cite{Bergman}, plays a crucial role. These dimensionless numbers guide us in selecting the appropriate heat transfer coefficient~$h_\mathrm{SS-LAr}$ for the stainless steel-to-LAr wall interaction. Considering the LAr flowing across a single unit tube within a hexagonally arranged tube bank at a design flow rate of $\SI{1}{slpm}$ (assuming a total flow of $\SI{1000}{slpm}$ across $\num{1000}$ tubes), our conditions satisfy $Gr/Re^2 \gg 1$ and $Ra \approx \num{e10}$. These values indicate a turbulent free convection regime. Based on reference~\cite{Bergman} (section 9.6.1), the Churchill and Chu correlation is applicable over the entire range of $Ra$:
\begin{align}
     h_\mathrm{SS-LAr} &= \frac{k_\mathrm{LAr}}{L} \left( 0.825 + \frac{0.387 Ra_\mathrm{LAr}^{1/6}}{[1 + (0.492/Pr)^{9/16}]^{8/27}} \right)^2 \quad.
\end{align}
The LAr properties are evaluated at the film's temperature $T_\mathrm{film,out} = (T_\mathrm{LAr,sat} + T_\mathrm{S,out}) / 2$.

Energy conservation dictates that the heat transfer must be equal across each layer. Consequently, the limiting factor for the overall heat transfer will be the mechanism with the lowest efficiency, in this case, the stainless steel wall to LAr transfer. We then search for a set of temperatures that fulfils the condition
\begin{equation}
   \dot  Q (T_\mathrm{GAr,sat}, T_\mathrm{S,in}, T_\mathrm{S,out},T_\mathrm{LAr,sat}) = \dot Q_\mathrm{GAr-SS} = \dot Q_\mathrm{SS} = \dot Q_\mathrm{SS-LAr} \quad.
\end{equation}
for a given operational pressure~$p_\mathrm{GAr}$ on the tube's inside, and a given pressure~$p_\mathrm{LAr}$ on the tube's outside. Subsequently, we vary the internal argon pressure~$p_\mathrm{GAr}$ and its corresponding temperature~$T_\mathrm{GAr,sat}$ to calculate the heat transfer per tube as a function of the pressure differential $(p_\mathrm{GAr} - p_\mathrm{LAr})$. The results of these calculations are compared to experimental data obtained with a 19"~tube heat exchanger in the following section.

\subsection{Argon-argon heat transfer measurement}
\label{appsubsec:MultiTubeSmallScaleTest}

\begin{figure}[t]
\centering
\begin{subfigure}[b]{0.6\textwidth}
\centering
\includegraphics*[height=0.2\textheight]{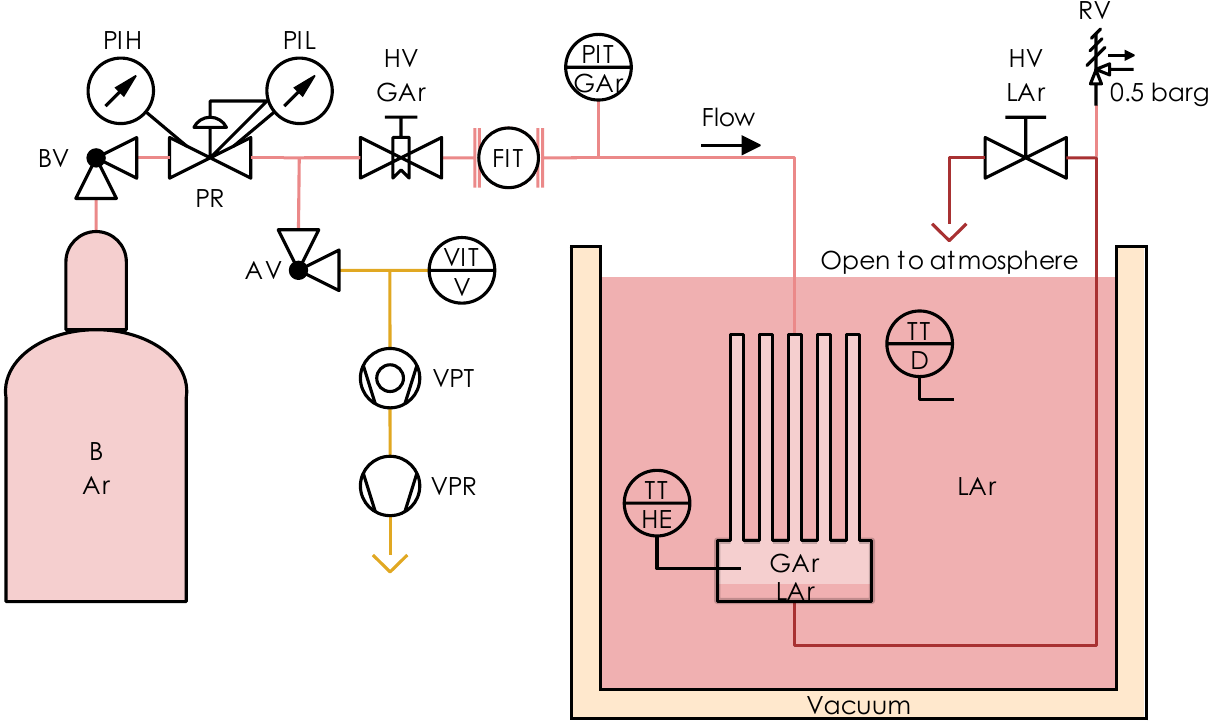}
\end{subfigure}
\quad
\begin{subfigure}[b]{0.36\textwidth}
\centering
\includegraphics*[height=0.2\textheight]{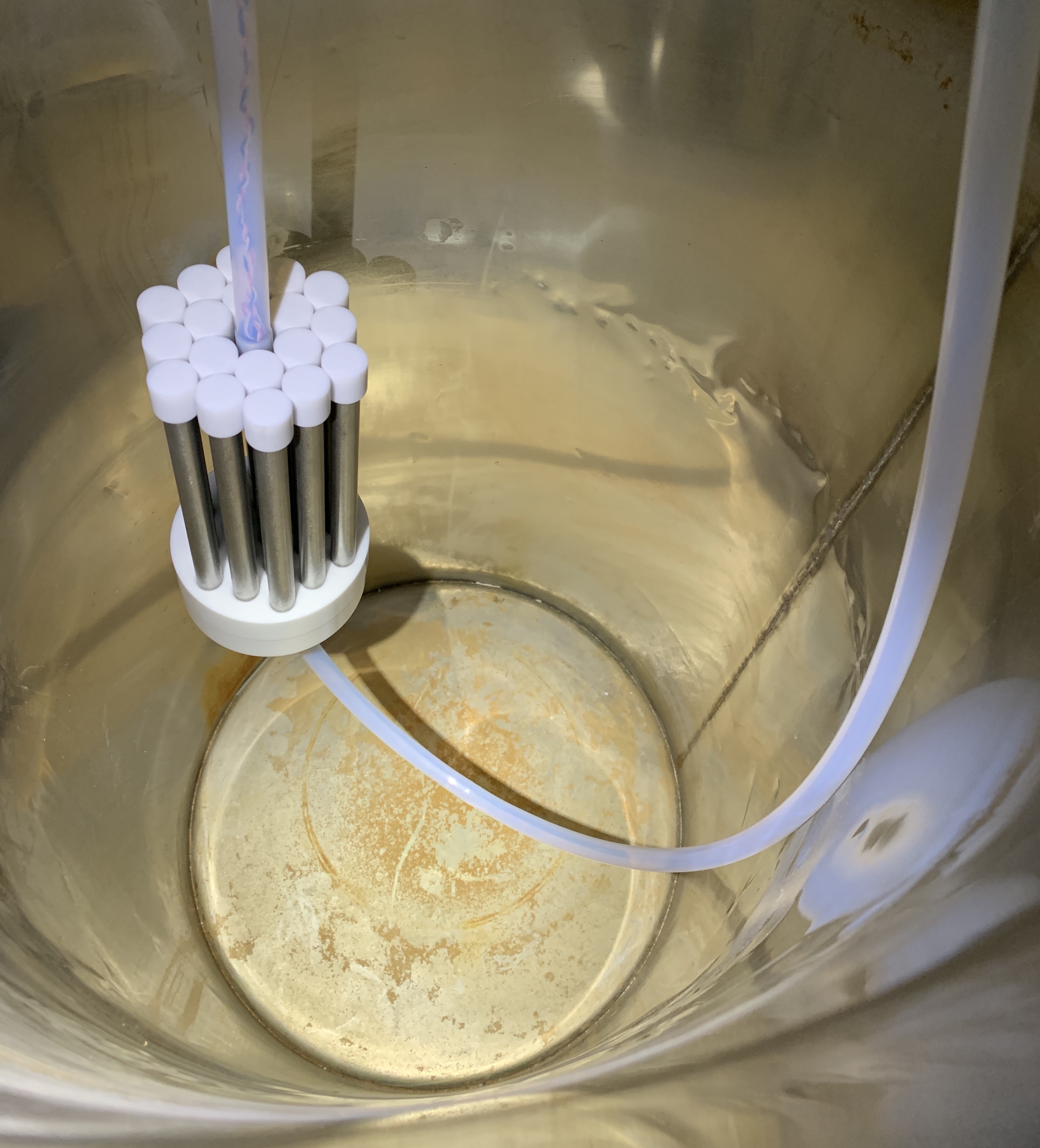}
\end{subfigure}
\caption[]{(Left): P\&ID of the small-scale heat transfer test with a 19"~tube heat exchanger. (Right): Picture of the 19"~tube heat exchanger inside the empty Dewar flask.}
\label{fig:TwoPhaseHESmallTest}
\end{figure}

In figure~\ref{fig:TwoPhaseHESmallTest}~(left) we show the P\&ID of the test setup. The heat exchanger is immersed in a LAr bath in an open Dewar flask. Room temperature GAr\footnote{In the DS"~20k cryogenic system, GAr arrives close to saturation temperature at the two-phase heat exchanger because it is located downstream of the LN$_{2}$-GAr condenser. This reduces the required total enthalpy change for liquefaction compared to the use of warm GAr.} from a gas cylinder was injected at fixed pressures into the heat exchanger. The condensation rate of GAr, and thus the rate of heat transfer, was measured through the gas flow rate into the heat exchanger. The condensed argon is collected in the reservoir below the tubes. The reservoir can be emptied via a drain by flushing the system with pressurised GAr. Pressure, flow and temperature data were recorded with a LabVIEW~\cite{labview} virtual instrument. The heat exchanger geometry consisted of 19~tubes arranged in the pattern described in section~\ref{subsec:CondenserBox}, see figure~\ref{fig:TwoPhaseHESmallTest}~(right). The white body of the heat exchanger and the connecting hoses are made from PTFE to minimise heat transfer via surfaces other than the ones of the tubes. The condensation rate from the hose was measured as a function of pressure difference and subtracted from the results. The pressure drop from the location of the pressure transducer to the heat exchanger at the maximum measured gas flow of $\SI{\sim 30}{slpm}$ over the short $\nicefrac{1}{2}\,\si{in}$"~piping is of the order of a few $\si{mbar}$ and thus negligible. Seals of the various components are made using press fits at room temperature between PTFE and stainless steel, which result in tight seals at cryogenic temperatures. This resulted in a vacuum level of the order of $\SI{e-5}{mbara}$ at cold and a negligible leak rate during an overpressure test with helium at $\SI{1315}{mbara}$, observed as tiny helium bubbles in the LAr. The helium pressure remained stable over minutes within the measurement accuracy. Different from the design used for the measurements with the test bed, in this configuration the $\SI{7}{in}$ long tubes were exposed to the LAr bath only at a length of $\SI{6}{in}$ because the ends were fitted into the PTFE body on one end and into PTFE end caps on the other.

\begin{figure}[t]
\centering
\begin{subfigure}[b]{0.48\textwidth}
\centering
\includegraphics*[width=1\textwidth]{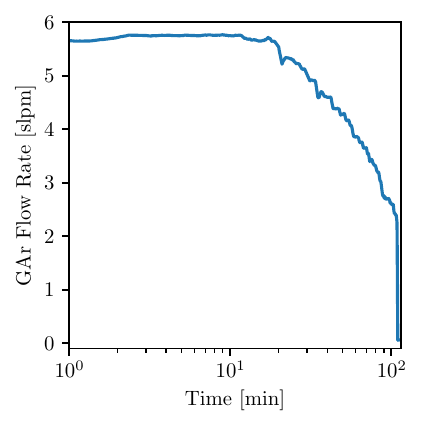}
\end{subfigure}
\quad
\begin{subfigure}[b]{0.48\textwidth}
\centering
\includegraphics*[width=1\textwidth]{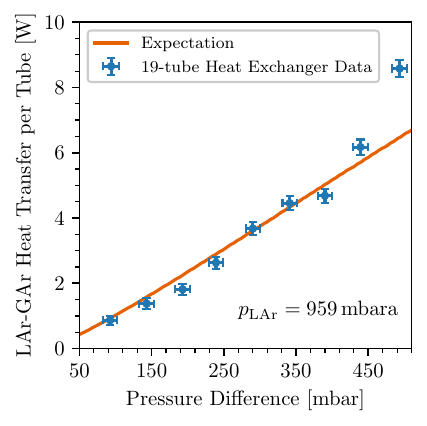}
\end{subfigure}
\caption[]{Data from the GAr-LAr heat transfer measurement with the 19"~tube heat exchanger. (Left): Observed GAr flow rate at a differential pressure of $\SI{143}{mbard}$ with respect to the ambient pressure of $\SI{959}{mbara}$. The raw data shown here is not corrected for the flow rate due to condensation on the PTFE hose. (Right): Heat transfer rate with GAr cooling and liquefaction as a function of overpressure of the supplied argon with respect to the ambient pressure. The orange curve is the expectation from modelling including turbulent free convection and film condensation. The deviation of the data from the expectation at high differential pressures is due to the onset of LAr boiling.}
\label{fig:TwoPhaseHESmallTestResults}
\end{figure}

The measured GAr flow time series at a differential pressure of $\SI{143}{mbard}$ with respect to ambient pressure is shown in figure~\ref{fig:TwoPhaseHESmallTestResults}~(left). We observed a constant and stable GAr flow while the PTFE bucket below the tubes was filled with condensed argon. As soon as the liquid level reached the tubes, the flow decreased approximately linearly with time until it stopped when the entire internal volume of the heat exchanger was filled. Note that the hydrostatic pressure due to the depth of the heat exchanger under the liquid surface does not contribute to the differential pressure since the temperature of the open LAr bath, which is the relevant quantity for the heat transfer rate, is defined by the ambient pressure.
From the product of the enthalpy change of the fluid and the constant mass flow rate observed during the initial phase, the heat transfer rate for the GAr cooling and liquefaction can be obtained. In figure~\ref{fig:TwoPhaseHESmallTestResults}~(right), the result is presented as a function of the differential between the pressure of the supplied GAr and the ambient pressure.
We observe a strong enhancement of the GAr-LAr heat transfer rate with differential pressure, which is driven by convective currents in the surrounding LAr.
The model predictions from the previous section closely match the measured data up to a differential pressure of approximately $\SI{450}{mbard}$. Beyond this point, the measured heat transfer exceeds the expected values. This discrepancy suggests a transition in the flow regime within the liquid argon. Video recordings from the experiment support this observation, indicating the onset of boiling at higher pressure differentials\footnote{As the final design of the DS"~20k heat exchanger will operate well below $\SI{500}{mbard}$, further modelling extending beyond this pressure range is not necessary.}.
While the GAr was injected at room temperature into the gas system of the installation, the GAr temperature inside the heat exchanger was lower due to cooling inside the PTFE hose during transfer to the heat exchanger. Thus, the enthalpy change used to calculate the heat transfer rate in figure~\ref{fig:TwoPhaseHESmallTestResults}~(right) is based on the GAr temperature measured inside the heat exchanger with TT~HE (cf.~figure~\ref{fig:TwoPhaseHESmallTest}~(left)).  

\begin{figure}[t]
\centering
\includegraphics[width=0.618\textwidth]{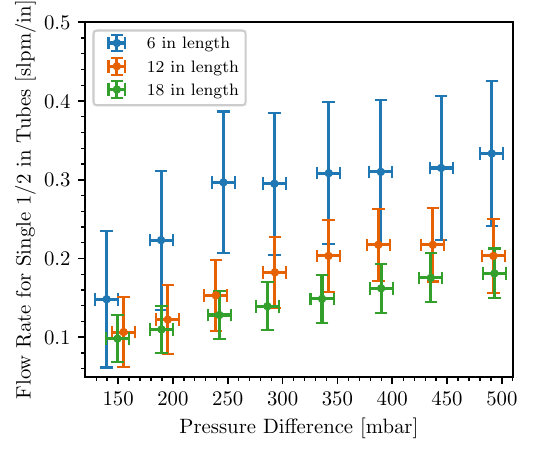}
\caption[]{Argon flow for single $\nicefrac{1}{2}\,\si{in}$~tubes of different lengths. The large flow errors are due to the accuracy of the flow meter set at $\SI{100}{slpm}$ full scale (model MV"~308 from Bronkhorst\textsuperscript{\textregistered}~\cite{Bronkhorst}).}
\label{fig:SinglePipeHEResults}
\end{figure}

The dependence of the heat transfer rate on the length of the tubes was evaluated in a separate test, similar to the one just described, with single isolated tubes of different lengths, see figure~\ref{fig:SinglePipeHEResults}. We find that the mean heat transfer rate in longer tubes is reduced due to the formation of a liquid film on the inner surface, which grows downwards and reduces the heat transfer coefficient between the GAr and the stainless steel tube.

\newpage
\bibliographystyle{jhep}
\bibliography{main}

\end{document}